\documentclass[%
reprint,
nofootinbib,
amsmath,amssymb,
aps,
]{revtex4-2}

\usepackage{graphicx}
\usepackage{dcolumn}
\usepackage{bm}

\usepackage{multirow}

\usepackage{amsmath,amssymb,amsfonts}
\usepackage{graphicx}
\usepackage{color}
\usepackage{appendix}
\usepackage{amsthm}
\usepackage{tikz}
\usepackage{youngtab}
\usepackage{mleftright}
\usepackage{url}
\usepackage[most]{tcolorbox}

\usepackage{url}
\usepackage{hyperref}

\usepackage{tikz}
\usetikzlibrary{shapes,snakes,positioning,automata,arrows.meta,calc,decorations.markings,math,arrows.meta}
\tikzstyle{vertex}=[circle, draw, inner sep=0pt, minimum size=6pt]
\usepackage{xcolor}
\usepackage{subfig}

\usepackage[export]{adjustbox}

\usepackage{geometry}
\newcommand{\nn}{\nonumber}

\def\a{\alpha}
\def\b{\beta}

\def\s{\sigma}

\def\t{\tau}

\def\p{\psi}

\def\P{\Psi}

\def\<{\langle}
\def\>{\rangle}

\def\mH{{\mathcal{H}}}

\def\ha{{\hat{a}}}

\def\tu{{\tilde{0}}}
\def\td{{\tilde{2}}}
\def\t2{{\tilde{1}}}

\usepackage[normalem]{ulem}

\newcommand\encircle[1]{%
	\tikz[baseline=(X.base)] 
	\node (X) [draw, shape=circle, inner sep=0] {\strut #1};}

\begin{document}

	\title{Shortcut to Multipartite Entanglement Generation: A Graph Approach to Boson Subtractions}

	\author{Seungbeom Chin}
	\email{sbthesy@skku.edu}
	\affiliation{International Centre for Theory of Quantum Technologies, University of Gda\'{n}sk, 80-308, Gda\'{n}sk, Poland \\ Department of Electrical and Computer Engineering, Sungkyunkwan University, Suwon 16419, Korea}

	\author{Yong-Su Kim}
	\affiliation{Center for Quantum Information, Korea Institute of Science and Technology (KIST), Seoul, 02792, Korea \\
		Division of Nano $\&$ Information Technology, KIST School, Korea University of Science and Technology, Seoul 02792, Korea}	
	
	\author{Marcin Karczewski}
	\email{marcin.karczewski@ug.edu.pl}
	\affiliation{International Centre for Theory of Quantum Technologies, University of Gda\'{n}sk, 80-308, Gda\'{n}sk, Poland}

	
	\begin{abstract}
	We propose a graph method for systematically searching for schemes that can generate multipartite entanglement in linear bosonic systems with heralding. While heralded entanglement generation offers more tolerable schemes for quantum tasks than postselected ones, it is generally more challenging to find appropriate circuits for multipartite systems. We show that our graph mapping from boson subtractions provides handy tactics to overcome the limitations in circuit designs.		We present a practical strategy to mitigate the limitation through the implementation of our graph technique. Our physical setup is based on the sculpting protocol, which utilizes an $ N$ spatially overlapped subtractions of single bosons to convert Fock states of evenly distributed bosons into entanglement.
	We have identified general schemes for qubit N-partite GHZ and W states, which are significantly more efficient than previous schemes. In addition, our scheme for generating the superposition of $N=3$ GHZ and W entangled states illustrates that our approach can be extended to derive more generalized forms of entangled states. Furthermore, we have found an N-partite GHZ state generation scheme for qudits, which requires substantially fewer particles than previous proposals.
	 These results demonstrate the power of our approach in discovering optimized solutions for the generation of intricate heralded entangled states. 
 As a proof of concept, we propose a linear optical scheme for the generation of the Bell state by heralding detections. 
We expect our method to serve as a promising tool in generating diverse entanglement.	
		
		
	\end{abstract}
	
	
	\maketitle

	\section{Introduction} \label{intro}
	
	Entanglement is an essential aspect of quantum information science, investigation of which has resulted in new fundamental understandings about the nature~\cite{horodecki2009quantum}. Practically, entanglement is recognized as a valuable resource in the field of quantum information processing, with potential applications in areas such as cryptography~\cite{gisin2002quantum,pirandola2020advances} and computing~\cite{knill2001scheme,preskill2018quantum}.
	To study and utilize entanglement, it is a prerequisite to find  reliable procedures  to construct entangled quantum systems. 
	One of the promising approaches to this task is to exploit the indistinguishability of quantum particles~\cite{dirac1981principles}.  Various works suggested theoretical and experimental  entanglement generation schemes based on the identicality of particles and postselection. Along these lines, Refs.~\cite{tichy2013entanglement, krenn2017entanglement, franco2018indistinguishability} showed that two spatially overlapped indistinguishable particles can carry bipartite entanglement. The quantitative relation of particle indistinguishability and spatial overlap to the bipartite entanglement was rigorously analyzed in Refs.~\cite{chin2019entanglement, nosrati2020robust, barros2020entangling}. For the case of multipartite entanglement, schemes for GHZ and W states with identical particles have been theoretically suggested~\cite{yurke1992einstein,yurke1992bell, blasiak2019entangling,bellomo2017n,kim2020efficient, krenn2017quantum,gu2019quantum,gu2019quantum2, blasiak2021efficient, chin2021graph} and experimented~\cite{erhard2018experimental,lee2022entangling}. Ref.~\cite{chin2021graph} presented a comprehensive graph-theoretic approach to embrace the schemes to generate the entanglement of identical particles in linear quantum networks (LQNs) with postselection.

	On the other hand, considering that
	the postselected schemes are highly sensitive to particle loss in circuits~\cite{gimeno2015three,pant2019percolation} and the multipartite correlations can be created by the postselection bias~\cite{glymour2016causal,gebhart2021genuine}, there have been several attempts to generate the entanglement of identical particles {\it{without postselection}}.  Specifically, heralded generation of entangled states of photons was  studied for bipartite~\cite{barz2010heralded,wagenknecht2010experimental,kim2009three,ra2015phase} and multipartite systems~\cite{papp2009characterization,schwartz2016deterministic,shi2013heralded,tavakoli2020autonomous,gubarev2020improved,le2021heralded}. 
	While  postselected schemes involve identifying desired outcomes after conducting operations, heralded schemes use real-time signals or measurements in ancillary spatial modes as ``heralds'' to sort out successful runs during the experiment (see, e.g., Ref.~\cite{gubarev2020improved} for a more detailed explanation).
	This property of heralded operations renders more tolerable schemes from photon loss~\cite{gimeno2015three}, however it is usually more challenging to find proper circuits to obtain the heralded entanglement of an arbitrary $N$-partite systems than the postselected ones~\cite{gubarev2020improved}.

	In this work, we introduce a systematic method to overcome the difficulty of obtaining heralded entanglement generation schemes for an arbitrary $N$-partite system. 
	Our method employs the \emph{sculpting protocol} introduced in Ref.~\cite{karczewski2019sculpting}, which generates an $N$-partite entangled state by
	applying $N$ single-boson subtraction operators (which we name the ``sculpting operator'') to a $2N$ boson initial state. By setting the initial state to have the even distribution of the bosons in different $2N$ states, the spatially overlapped sculpting operation (i..e, a single-boson subtraction operator $\hat{A}^{(l)}$ is a summation of subtractions on different spatial modes) generates the $N$-partite entanglement. And various sculpting operators result in different entangled states. Since linear bosonic systems with heralding detectors can realize boson subtraction operators~\cite{kim2008scheme,zavatta2009experimental,parigi2007probing,ourjoumtsev2006generating}, we can design $N$-partite genuine entanglement generation schemes  with this theoretical process.

\begin{center}
		\begin{figure}
			\centering
			\includegraphics[width=.47\textwidth]{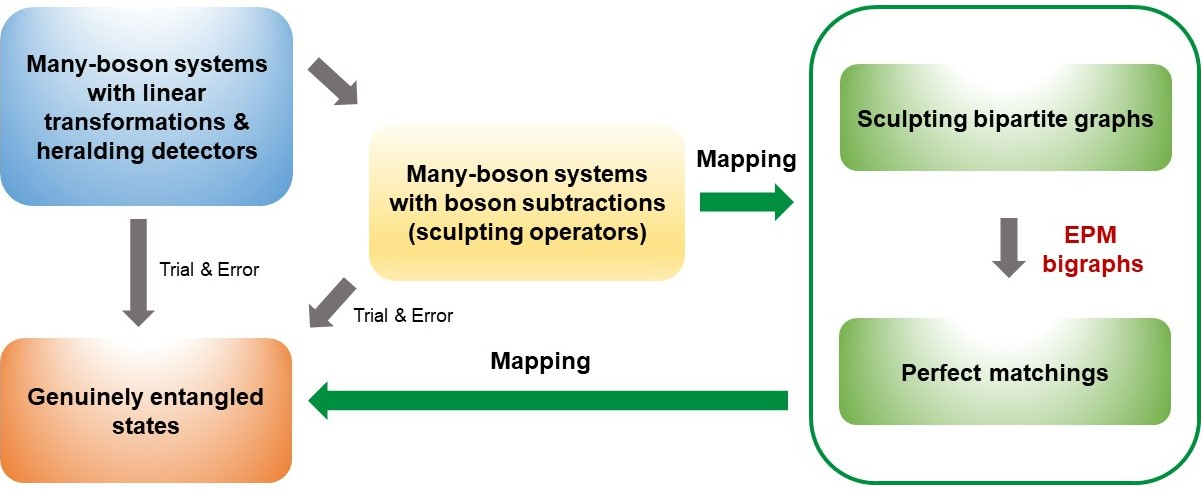}
			\caption{Previous research on generating entangled states has primarily relied on trial and error methods via the direct route depicted by the blue to red boxes, or by taking a detour through the yellow box. However, we present a more systematic approach to designing circuits for various entangled states by mapping the elements of many-boson systems onto graphs, as illustrated by the routes of the green arrows.  }
			\label{graph}
		\end{figure}
		
	\end{center}

	In the sculpting protocol, the difficulty of designing a circuit for an $N$-partite entangled state is translated to the difficulty of finding a suitable sculpting operator to ``chisel'' the state. However, former research on the sculpting protocol~\cite{karczewski2019sculpting,zaw2022sculpting} provides only proof-of-concept schemes demonstrating the method and lacks any systematic way of linking the features of sculpting operators to the expected final states. 
	Our work shows that \emph{a graph picture of the sculpting protocol provides an straightforward strategy of finding appropriate sculpting operators for entanglement, hence appropriate heralding schemes themselves.}
	We map multi-boson systems with sculpting operators into bipartite graphs (bigraphs), for which we develop techniques to understand key properties of the entanglement generation process. Our list of correspondence relations between sculpting protocols and graphs is a variation of that given in Ref.~\cite{chin2019entanglement}, which provided a systematic method to analyze and design LQNs for obtaining entanglement with postselection.
	In this graph picture, we have found a special type of bigraphs, which we name \emph{effective perfect matching (EPM) bigraphs}. These bigraphs are highly useful because they can directly correspond to sculpting operators that generate entanglement. Fig.~\ref{graph} describes our graph strategy to search for genuinely entangled states.
	
	With our graph-theoretic approach, we present sculpting operators that generate qubit $N$-partite GHZ and W states, and an $N=3$ Type 5 entangled state (the superposition of GHZ and W states~\cite{acin2000generalized}). The GHZ and W schemes are significantly more efficient than those given in Ref.~\cite{karczewski2019sculpting}. And contrary to the schemes in Refs.~\cite{gimeno2015three,gubarev2020improved}, they work for arbitrary number of parties.
	In addition,  our $N=3$ Type 5 entangled state generation scheme illustrates that our approach can be extended to find more generalized forms of entangled states.
	
	To top it off, by generalizing the bigraph used to obtain qudit GHZ states, we also present a $qudit$ $N$-partite GHZ state generation scheme with $dN$ bosons. To our knowledge, our scheme requires much less bosons than any known heralded schemes as in Ref.~\cite{paesani2021scheme}.
	Our theoretical schemes can be realized in any many-boson system, e.g.,  linear optical systems with polarization qubit encoding and heralded detections. 
	These outcomes showcase the effectiveness of our method in finding simple solutions for the generation of intricate entangled states.
	
	Our work is organized as follows: Sec.~\ref{sculpting protocol} reviews the sculpting protocol introduced in Ref.~\cite{karczewski2019sculpting}. Sec.~\ref{graph picture} explains our dictionary of mapping the sculpting protocol to bigraphs. We also show that the perfect matchings (PMs) of bigraphs determine the final state after the sculpting operation. Sec.~\ref{qubit_graphs} gives sculpting operators that generate qubit $N$-partite GHZ and W states. Using the qubit GHZ generation graph, Sec.~\ref{qudit} presents a qudit GHZ state generation scheme. 
	Sec.~\ref{experiments} explains how linear optical systems with polarization qubit encoding and heralding detectors can build our sculpting schemes. To showcase our method, we present a simple Bell state generation example.  Sec.~\ref{discussions} summarizes the significance of our results and discusses possible follow-up researches.
	
	 For the video summary of this paper, go to \url{https://youtu.be/iQ4aWQJuRZI}.
	
	\section{Sculpting protocol for qubit entanglement}\label{sculpting protocol}
	
	In this section, we formalize the sculpting protocol~\cite{karczewski2019sculpting} that converts the boson identity
	into entanglement. While $N$-partite entangled state was constructed in Ref.~\cite{karczewski2019sculpting} based on $2N$ modes with the dual-rail qubit encoding, we re-explain it based on $N$ spatial modes and consider the qubit state as a two-dimensional internal degree of freedom of bosons. This way of expression not only embraces the dual-rail encoding, but also provides a more intuitive description of qubit states in the system.   
	
\begin{figure}
		\centering
		\includegraphics[width=7.5cm]{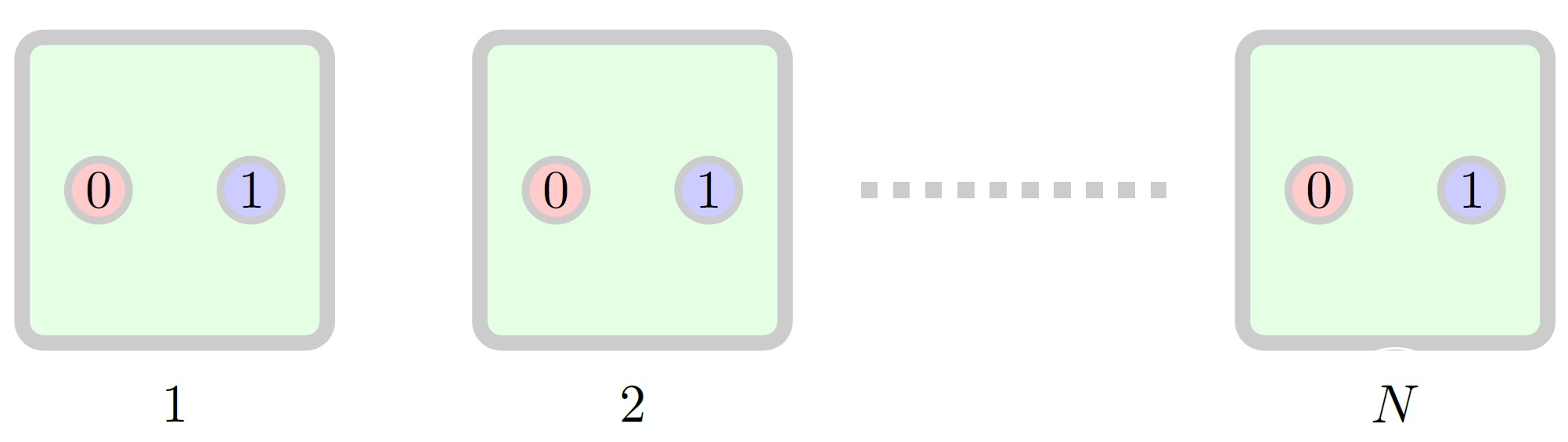}
		\caption{The initial state  $|Sym_{N}\>$ of $2N$ bosons in $N$ spatial modes. Each spatial mode has two bosons, one in the internal state $|0\>$ and the other in $|1\>$. }
		\label{fig_initial}
	\end{figure}

	Since in our setup each boson in $j$th spatial mode ($j\in \{1,2,\cdots, N\}$) has a two-dimensional internal degree of freedom $s$ $(\in \{0,1\})$, boson creation and annihilation operators are denoted as $\ha_{j,s}^\dagger$ and $\ha_{j,s}$ respectively.
	Then, as an input state, we distribute $2N$ bosons into $N$ spatial modes so that each spatial mode has two bosons with orthogonal internal states ($0$ and $1$, see Fig.~\ref{fig_initial}). Therefore, the initial state is given by 
	\begin{align}\label{initial}
		|Sym_{N}\> &\equiv \ha^\dagger_{1,0}\ha^\dagger_{1,1} \ha^\dagger_{2,0}\ha^\dagger_{2,1} \cdots 
		\ha^\dagger_{N,0}\ha^\dagger_{N,1}|vac\>  \nn \\
		&= \prod_{j=1}^N(\ha^\dagger_{j,0}\ha^\dagger_{j,1})|vac\>. 
	\end{align} 
	
	Following the former works~\cite{karczewski2019sculpting,zaw2022sculpting}, we call it the \emph{maximally symmetric state} of $2N$ bosons. 
	Rewriting $|Sym_N\>$ in the mode occupation representation as 
	\begin{align}
		|Sym_N\> = |(1,1),(1,1),\cdots (1,1)\>
	\end{align} and the particle number distribution in the $2N$ states as a vector, we see that $|Sym_N\>$ is majorized by all the other Fock states of $2N$ bosons (i.e., since each mode has one particle, the particle number distribution vector of $|Sym_N\>$ is written as $\vec{n}_{Sym} = (1,1,\cdots, 1)$. This vector is majorized by any particle number distribution vector of $2N$ dimension. See Ref.~\cite{chin2019majorization}, Sec. II for the rigorous definitions and analyses). Several research papers showed that this kind of state is very resourceful in many quantum computation protocols~\cite{latorre2002majorization,orus2004systematic,chin2019majorization}. 

	What we need to obtain in the sculpting protocol is a state in which each spatial mode has one boson whose internal state corresponds to the qubit state. For such a final state, we need to annihilate $N$ single-bosons from the initial state $|Sym_N\>$. The $N$ single-boson subtraction operators, which we name the \emph{sculpting operator}, is expressed in the most general form as 
	\begin{align}\label{annihilation}
		&	\prod_{l=1}^N\sum_{j=1}^N(k^{(l)}_{j ,0}\ha_{j,0} + k^{(l)}_{j,1}\ha_{j,1} ) \nn \\
		&\equiv \prod_{l=1}^N\hat{A}^{(l)} \equiv \hat{A}_N \nn \\
		& \quad (k^{(l)}_{j,s} \in \mathbb{C}~\textrm{and}~  \sum_{j,s}|k^{(l)}_{j,s}|^2 =1).
	\end{align}
	We see that the one-boson subtraction operator $\hat{A}^{(l)}$ can superpose among different spatial modes. 	 Such an operation has been implemented probabilistically in several bosonic experimental  setups~\cite{kim2008scheme,zavatta2009experimental,parigi2007probing,ourjoumtsev2006generating}.
	
	For a later convenience, we rewrite $\hat{A}^{(l)}$ as
	\begin{align}\label{annihilation}
		\hat{A}^{(l)} &= \sum_{j=1}^N(k^{(l)}_{j ,0}\ha_{j,0} + k^{(l)}_{j,1}\ha_{j,1} )   \nn \\
		&\equiv   \sum_{j=1}^N\a_{j}^{(l)}\ha_{j,\p_{j}^{(l)}}
	\end{align}
	where $\a_j^{(l)}\in \mathbb{C}$ with $\sum_{j}|\a^{(l)}_j|^2 =1$ and $|\p_{j}^{(l)}\>$ a normalized qubit state.

	Applying the sculpting operator $\hat{A}_N$ to the initial state $|Sym_{N}\>$, we obtain the final  state,
	\begin{align}\label{final_state}
		|\P\>_{fin} = \hat{A}_N|Sym_{N}\>. 
	\end{align}
	The correlation of $|Sym_N\>$  from the indistinguishability of bosons  and the spatial overlap among bosons that comes from $\hat{A}_N$  establish the entanglement property of $|\P\>_{fin}$. 
 Therefore, we can state that \emph{the indistinguishability of identical particles and spatial overlap are two essential elements for the entanglement generation with sculpting protocol,} as in postselected schemes~\cite{franco2018indistinguishability, paunkovic2004role, barros2020entangling,lee2022entangling}.
	
	There is an essential restriction that we need to impose on a sculpting operator $\hat{A}_N$: it must be a sum of operators that annihilates one particle per spatial mode so that the final total state $|\P\>_{fin}$ must consist of states with one particle per spatial mode. Then the internal state of one particle encodes the qubit information.
		This also means that any term of $\hat{A}_N$ that subtracts both particles from a given spatial mode must vanish. Otherwise, as the total number of subtracted particles is fixed, there remains a term with two particles in the same spatial mode in $|\Psi\>$. We call this restriction the \emph{no-bunching restriction}. We can find such an $\hat{A}_N$ that satisfies the restriction by controlling the probability amplitudes of it. 
	
	For the simplest $N=2$ example,  $\hat{A}_2$ is written as
		\begin{align}
			\hat{A}_2 &= \hat{A}^{(1)}_2\hat{A}^{(2)}_2 \nn \\
			&=(\a_1\ha_{1\p_1} +\a_2\ha_{2\p_2})(\b_2\ha_{1\phi_1} + \b_2\ha_{2\phi_2})
		\end{align}	
		and the final state is given by
		\begin{align}
			|\Psi\>_{fin} &= \hat{A}_2|Sym_2\> \nn \\
			& = (\a_1\b_2\ha_{1\p_1}\ha_{2\phi_2} + \a_2\b_1\ha_{2\p_2}\ha_{1\phi_1})|Sym_2\> \nn \\
			&~~~+\a_1\b_1\ha_{1\p_1}\ha_{1\phi_1}|Sym_2\>\nn \\
			&~~~+  \a_2\b_2\ha_{2\p_2}\ha_{2\phi_2}|Sym_2\>.
		\end{align}
		In the above equation, $\a_1\b_1\ha_{1\p_1}\ha_{1\phi_1}|Sym_2\>$ must vanish, otherwise the term has two particles in the second spatial mode which violates the no-bunching condition. For the same reason,  $\a_2\b_2\ha_{2\p_2}\ha_{2\phi_2}|Sym_2\>$ also must vanish. 
	
	In the dual rail encoding setup~\cite{karczewski2019sculpting,zaw2022sculpting}, the no-bunching condition appears as a seemingly different form. In that setup, repetitive annihilations on the same spatial mode naturally vanish. However, since two spatial modes combine to constitute one subsystem for the case, valid final states are only restricted to those with one boson per two spatial modes. This exactly corresponds to the no-bunching restriction in our setup.

	All things considered, we summarize the sculpting protocol as follows:
	\begin{tcolorbox}[enhanced jigsaw,colback=black!10!white,boxrule=0pt,arc=0pt]	
		\textbf{Sculpting protocol}
		\begin{enumerate}
			\item Initial state: We prepare the maximally symmetric state $|Sym_{N}\>$ of $2N$ bosons, i.e., each boson has different states (either spatial or internal) with each other as Eq.~\eqref{initial} (see Fig.~\ref{fig_initial}). 
			\item Operation: We apply the sculpting  operator $\hat{A}_N$ of the form~\eqref{annihilation} to the initial state $|Sym_{N}\>$. \emph{The sculpting process must satisfy the no-bunching condition.} 
			\item Final state: The final state can be fully separable, partially separable, or genuinely entangled.
		\end{enumerate}
	\end{tcolorbox}	
	It is worth mentioning that the degree of entanglement for an $N$-partite pure state can be categorized into three classes: fully separable, partially separable, and genuinely entangled~\cite{walter2016multipartite} (see Ref.~\cite{chin2021graph}, 3.1 for a quick summary of these concepts). A state $|\p\>$ ($\in \mH = \otimes_{j=1}^N \mH_i$) is fully separable if it can be written as $|\p\> = |\p_1\>\otimes |\p_2\>\otimes \cdots \otimes |\p_N\>$ where $|\p_j\>\in \mathcal{H}_j$ for all $j=1,2,\cdots ,N$. It is genuinely entangled if 
		it cannot be separable
		under any bipartition of $\mathcal{H}$.	
		It is partially separable when it is neither fully separable nor genuinely entangled. The states we target in the work are genuinely entangled states such as GHZ and W states. 
	
	Another crucial remark is that a sculpting operator can generate an $N$-partite entangled state with $K$ ancillary spatial modes. Then the above sculpting protocol can be slightly varied to an $(N+K)$-partite intial state with the sculpting operator $\hat{A}_{N+K}$. We can see such cases in Sec.~\ref{W_qubit_N} and~\ref{Type_5}.

	Most of the technical difficulty to find $\hat{A}_N$ for a specific entanglement state comes from Step 2,  for it is critical to control the probability amplitudes so that the sculpting operator satisfies the no-bunching restriction. There have been no systematic technique to find a suitable $\hat{A}_N$ that simultaneously satisfies the no-bunching condition and generates non-trivial entanglement state~\cite{karczewski2019sculpting}. As we will explain in the following sections, our \emph{graph techniques} facilitate a powerful tool to overcome this limitation.
	

	$ $\\	
	
	\section{Graph picture of boson systems with sculpting operators}\label{graph picture}
	
	In this section, we present a list of correspondence relations between the fundamental elements of the sculpting protocol and those of graphs. With the mapping, we can replace  key physical properties and restrictions  on the sculpting operators with those on graphs, which renders a handy guideline to the operator-finding process for genuinely entangled states. 
	
	Ref.~\cite{chin2021graph} proposed a list of correspondence relations between linear quantum networks (LQNs) and graphs for providing a systematic method to analyze and design networks for obtaining entanglement without postselection. Since our sculpting protocol also consists of linear transformations of boson subtraction operators, a similar graph mapping dictionary can be imposed to find a suitable $\hat{A}_N$ that generates genuine entanglement. Indeed, we can map spatially overlapped subtraction operators into graph elements with a variation of the correspondence relations in Ref.~\cite{chin2021graph}, which leads to a practical graph-theoretic method to analyze our system.

	The correspondence relations of elements between bosonic systems with sculpting operators and bigraphs can be enumerated as follows:
	
	\begin{center} 
		\begin{table}[h]
			\begin{tabular}{|l|l|l|}
				\hline
				\textbf{Boson systems} & \textbf{Bipartite Graph}  \\
				\textbf{with sculpting operator} & ~~~~~\textbf{$G_b =(U\cup V, E)$} \\
				\hline\hline 
				Spatial modes & Labelled vertices $\in$ $U$  \\ \hline 
				$\hat{A}^{(l)}$  ($l \in \{1,2,\cdots, N\}$) & Unlabelled vertices $\in$  $V$ \\ \hline 
				Spatial distributions of $\hat{A}^{(l)}$  & Edges  $\in$ $E$ \\ \hline 
				Probability amplitude $\a_j^{(l)} $ & Edge weight $\a_j^{(l)} $ \\ \hline 
				Internal state $\p_j^{(l)} $ & Edge weight $\p_j^{(l)}$  \\
				\hline 
			\end{tabular}
			\label{dict}
			\caption{Correspondence relations of a sculpting operator to a sculpting bigraph}
		\end{table} 
	\end{center} 
	In the above table, $\a_j^{(l)}$ and $\p_j^{(l)}$ are defined as in Eq.~\eqref{annihilation}. 
	A brief glossary in graph theory can be found in Ref.~\cite{chin2021graph}, Appendix A.

	In our graph picture, a subtraction operator$\hat{A}^{(l)} = \sum_{j=1}^N\a_{j}^{(l)}\ha_{j,\p_{j}^{(l)}} $ is denoted as an unlabelled vertex in $V$.
		Dynamical variables specifying the operator, such as spatial distributions and internal states, are encoded  as weighted edges connecting $V$ to labelled vertices in $U$.
	Below, unlabelled and labelled vertices are drawn as dots ($\bullet$) and circles (\encircle{$j$}) respectively. The array of subtraction operators (dots) are on the right hand side of the array of spatial modes (circles). We can consider a more comprehensive mapping including creation operators, which is given in Appendix~\ref{comprehensive list}. However, Table~I suffices to analyze the crucial properties of sculpting operators for generating entanglement.
	
	
	As a proof of concept, we analyze the simplest $N=2$ example with $\hat{A}_{2} = \hat{A}^{(1)}\hat{A}^{(2)}$.
	Let us write $\hat{A}^{(1)}$ and $\hat{A}^{(2)}$  as
	\begin{align}
		\hat{A}^{(1)}=& \a_1\ha_{1\p_1} + \a_{2}\ha_{2\p_2}, \nn \\
		\hat{A}^{(2)}=& \b_1\ha_{1\phi_1} + \b_{2}\ha_{2\phi_2}.
	\end{align}
	Then $\hat{A}^{(1)}$ applied to the system is mapped to a bipartite graph (bigraph)
	\begin{align} 
		\hat{A}^{(1)}=~	\begin{tikzpicture}[baseline={([yshift=-.5ex]current bounding box.center)}]
			\node[circle,draw,minimum size=0.6cm] (1) at (0,2) {$1$};
			\node[circle,draw,minimum size=0.6cm] (2) at (0,0) {$2$}; 
			\node[fill, vertex] (v) at (2,1) { };
			\path[line width = 0.8pt] (1) edge   node {$(\a_1, \p_1)$ } (v);
			\path[line width = 0.8pt] (2) edgenode {$(\a_{2},\p_2)$ } (v);
		\end{tikzpicture}.
	\end{align}
	Now, by applying $\hat{A}^{(2)}$, the total sculpting operator $\hat{A}_2$ corresponds to
	\begin{align}\label{N=2_A}
		\hat{A}_{2}=\hat{A}^{(2)}\hat{A}^{(1)}  =~ 
		\begin{tikzpicture}[baseline={([yshift=-.5ex]current bounding box.center)}]
			\node[circle,draw,minimum size=0.6cm] (1) at (0,2.4) {$1$};
			\node[circle,draw,minimum size=0.6cm] (2) at (0,0) {$2$}; 
			\node[fill, vertex] (v) at (3,2.4) { };
			\path[line width = 0.8pt] (1) edge node[above] {$(\a_1, \p_1)$ } (v);
			\path[line width = 0.8pt] (2) edge node[near end] {$(\a_2, \p_2)$} (v);
			\node[fill, vertex] (u) at (3,0) { };
			\path[line width = 0.8pt ] (1) edge   node[near end] {$(\b_1, \phi_1)$} (u);
			\path[line width = 0.8pt] (2) edge node[below] {$(\b_2, \phi_2)$ } (u);
		\end{tikzpicture}.
	\end{align} Note that the physical system is invariant under the exchange of two unlabelled vertices (dots), i.e., $\hat{A}_{2}$ can be also expressed as
	\begin{align}
		\hat{A}_{2} =\hat{A}^{(1)}\hat{A}^{(2)}  =~ 
		\begin{tikzpicture}[baseline={([yshift=-.5ex]current bounding box.center)}]
			\node[circle,draw,minimum size=0.6cm] (1) at (0,2.4) {$1$};
			\node[circle,draw,minimum size=0.6cm] (2) at (0,0) {$2$}; 
			\node[fill, vertex] (v) at (3,2.4) { };
			\path[line width = 0.8pt] (1) edge node [near end]  {$(\a_1, \p_1)$ } (u);
			\path[line width = 0.8pt] (2) edge node[below] {$(\a_2, \p_2)$} (u);
			\node[fill, vertex] (u) at (3,0) { };
			\path[line width = 0.8pt ] (1) edge   node[above] {$(\b_1, \phi_1)$} (v);
			\path[line width = 0.8pt] (2) edge node [near end] {$(\b_2, \phi_2)$ } (v);
		\end{tikzpicture}.
	\end{align}
	This represents nothing but the commutation relation $[\hat{A}^{(1)}, \hat{A}^{(2)}]=0$.
	
	When $\hat{A}_2$ is expanded as
	\begin{align}\label{A_2_expansion}
		\hat{A}_{2}&=\hat{A}^{(1)}\hat{A}^{(2)}  \nn \\
		&= \a_1\b_1\ha_{1\p_1}\ha_{1\phi_1} +  \a_1\b_2\ha_{1\p_1}\ha_{2\phi_2} \nn \\
		&~~~ + \a_2\b_1\ha_{2\p_2}\ha_{1\phi_1} + \a_2\b_2\ha_{2\p_2}\ha_{2\phi_2},
	\end{align}
	each term corresponds to a possible \emph{collective path} (a possible connection of dots to circles in which each dot is uniquely connected to one circle) of the annihilation operators, e.g., the bigraph~\eqref{N=2_A} has four possibilities for two annihilation operators to be applied to the spatial modes. Therefore, the expansion of $\hat{A}_{2}$~\eqref{A_2_expansion} is expressed with collective  paths as
	\begingroup
	\begin{align}\label{N=2_A_expansion}
		&\hat{A}_{2} \nn\\
		\phantom{sdf}\nn \\
		=~&
		\begin{tikzpicture}[baseline={([yshift=-.5ex]current bounding box.center)}]
			\node[circle,draw,minimum size=0.6cm] (1) at (0,2) {$1$};
			\node[circle,draw,minimum size=0.6cm] (2) at (0,0) {$2$}; 
			\node[fill, vertex] (v) at (2.5,2) { };
			\path[line width = 0.8pt] (1) edge node[below]  {$(\a_1, \p_1)$ } (v);
			\node[fill, vertex] (u) at (2.5,0) { };
			\path[line width = 0.8pt ] (1) edge   node[near end] {$(\b_1, \phi_1)$} (u);	
		\end{tikzpicture} +
		\begin{tikzpicture}[baseline={([yshift=-.5ex]current bounding box.center)}]
			\node[circle,draw,minimum size=0.6cm] (1) at (0,2) {$1$};
			\node[circle,draw,minimum size=0.6cm] (2) at (0,0) {$2$}; 
			\node[fill, vertex] (v) at (2.5,2) { };
			\path[line width = 0.8pt] (2) edge node[near end] {$(\a_2, \p_2)$} (v);
			\node[fill, vertex] (u) at (2.5,0) { };
			\path[line width = 0.8pt] (2) edge node[above] {$(\b_2, \phi_2)$ } (u);
		\end{tikzpicture}\nn \\
		\phantom{sdfdfd}\nn \\
		+&
		\begin{tikzpicture}[baseline={([yshift=-.5ex]current bounding box.center)}]
			\node[circle,draw,minimum size=0.6cm] (1) at (0,2) {$1$};
			\node[circle,draw,minimum size=0.6cm] (2) at (0,0) {$2$}; 
			\node[fill, vertex] (v) at (2.5,2) { };
			\path[line width = 0.8pt] (1) edge node[below] {$(\a_1, \p_1)$ } (v);
			\node[fill, vertex] (u) at (2.5,0) { };
			\path[line width = 0.8pt] (2) edge node[above] {$(\b_2, \phi_2)$ } (u);
		\end{tikzpicture} +
		\begin{tikzpicture}[baseline={([yshift=-.5ex]current bounding box.center)}]
			\node[circle,draw,minimum size=0.6cm] (1) at (0,2) {$1$};
			\node[circle,draw,minimum size=0.6cm] (2) at (0,0) {$2$}; 
			\node[fill, vertex] (v) at (2.5,2) { };
			\path[line width = 0.8pt] (2) edge node[near end] {$(\a_2, \p_2)$} (v);
			\node[fill, vertex] (u) at (2.5,0) { };
			\path[line width = 0.8pt ] (1) edge   node[near end] {$(\b_1, \phi_1)$} (u);
		\end{tikzpicture},
	\end{align}
	\endgroup
	i.e., $\hat{A}_2$ is a superposition of the above four collective paths. 
	
	For the sculpting operator $\hat{A}_{2}$ to obey the no-bunching condition, we must set the amplitudes so that the first two  collective paths in~\eqref{N=2_A_expansion} vanish when they are applied to $|Sym_{2}\> = \ha^\dagger_{1,0}\ha^\dagger_{1,1}\ha^\dagger_{2,0}\ha^\dagger_{2,1}|vac\>$. We can achieve such a sculpting operator by setting
	\begin{align}\label{N=2_amplitudes}
		& \a_j=\b_j=\frac{1}{\sqrt{2}}~~(j\in\{ 1,2\}) \nn \\
		& |\p_1\>=|\phi_2\> =\frac{1}{\sqrt{2}}(|0\> +|1\>) \equiv |+\>, \nn \\
		& |\p_2\>=|\phi_1\> = \frac{1}{\sqrt{2}}(|0\> -|1\> )\equiv |-\>. 
	\end{align}   
	Then it is direct to check that
	\begin{align}\label{N=2_A_final}
		&\hat{A}_{2}|Sym_{2}\>  \nn \\
		&=\Bigg(~
		\begin{tikzpicture}[baseline={([yshift=-.5ex]current bounding box.center)}]
			\node[circle,draw,minimum size=0.6cm] (1) at (0,1.5) {$1$};
			\node[circle,draw,minimum size=0.6cm] (2) at (0,0) {$2$}; 
			\node[fill, vertex] (v) at (1.8,1.5) { };
			\path[line width = 0.8pt] (1) edge node [above] {$(\frac{1}{\sqrt{2}}, +)$ } (v);
			\node[fill, vertex] (u) at (1.8,0) { };
			\path[line width = 0.8pt] (2) edge node [below] {$(\frac{1}{\sqrt{2}}, +)$ } (u);
		\end{tikzpicture} + \begin{tikzpicture}[baseline={([yshift=-.5ex]current bounding box.center)}]
			\node[circle,draw,minimum size=0.6cm] (1) at (0,1.5) {$1$};
			\node[circle,draw,minimum size=0.6cm] (2) at (0,0) {$2$}; 
			\node[fill, vertex] (v) at (1.8,1.5) { };
			\path[line width = 0.8pt] (2) edge node[near end] {$(\frac{1}{\sqrt{2}}, -)$} (v);
			\node[fill, vertex] (u) at (1.8,0) { };
			\path[line width = 0.8pt ] (1) edge   node[near end] {$(\frac{1}{\sqrt{2}}, -)$} (u);
		\end{tikzpicture}~\Bigg)|Sym_{2}\> \nn \\
		\phantom{text}\nn \\ 
		&= \frac{1}{2}(\ha_{1+}^\dagger\ha_{2+}^\dagger + \ha_{1-}^\dagger\ha_{2-}^\dagger)|vac\>,
	\end{align}
	i.e., by fixing amplitudes as Eq.~\eqref{N=2_amplitudes},  we obtain a Bell state as the final state.

	
	From the above $N=2$ example, we can understand the role of the no-bunching condition in the graph picture. Since the bigraph expression of $\hat{A}_N$ such as \eqref{N=2_A} is expanded with a summation of all the possible collective  paths of annihilation operators as Eq.~\eqref{N=2_A_expansion}, \emph{we have to control the complex weights of the edges so that any collective path with more than two edges in the same circle does not contribute to the final state.} This property can be understood with the concept of \emph{perfect matchings} (PMs), which are independent sets of edges in which every
		vertex of U is connected to exactly one vertex of
		V (see Ref.~\cite{chin2021graph}, Appendix A), as follows:
	
	\begin{tcolorbox}[enhanced jigsaw,colback=black!10!white,boxrule=0pt,arc=0pt]
		\textbf{Property 1.} For a specific sculpting operator $\hat{A}_N$, the final state $|\P\>_{fin} = \hat{A}_N|Sym_N\>$ must be fully determined by \emph{the addition of the  perfect matchings} (PMs) of the bigraph corresponding to $\hat{A}_N$.
	\end{tcolorbox}
	Indeed, we see that the two collective paths in Eq.~\eqref{N=2_A_final} are the two perfect matchings of the bigraph~\eqref{N=2_A}. The above property is useful for understanding given sculpting operators in several aspects, which we explain in Appendix~\ref{strategy} with a general \emph{sculpting-operator-finding strategy} based on this property. From now on, a bigraph that corresponds to a sculpting operator is called a \emph{sculpting bigraph}. 

	\section{Qubit entanglement: GHZ and W  states}\label{qubit_graphs}
	
	In this section, we present sculpting operators  that generate  qubit $N$-partite GHZ state, $N$-partite W state and a superposition of $N=3$ GHZ and W states, using Property 1. Our operator solutions are more efficient and more feasible to construct in many-boson systems than those given in Ref.~\cite{karczewski2019sculpting}, especially for the W state case.
	
	To find the sculpting operators for those entangled  states, we define a specially convenient type of bigraphs, which we dub \emph{effective PM bigraphs} (EPM). We restrict our attention to sculpting bigraphs whose edge weights of internal states are only among $\{|0\>,|1\>, |+\>,|-\> \}$ with $|\pm\>\equiv \frac{1}{\sqrt{2}}(|0\> \pm  |1\>)$.
	
	Among the creation and annihilation operators in the above basis, we can easily see the following identity
	\begin{align}\label{qubit_identities}
		\forall j & \in\{1,2,\cdots, N\}, \nn \\
		&\ha_{j,\pm}\ha^\dagger_{j,0}\ha^\dagger_{j,1}|vac\> = \pm\ha^\dagger_{j,\pm}|vac\>,
	\end{align} 
	holds, which directly results in the following identities:
	\begin{align}\label{qubit_identities_2}		&\ha_{j,+}\ha_{j,-}\ha^\dagger_{j,0}\ha^\dagger_{j,1}|vac\> = 0, \nn \\
		& \ha_{j,0}^{n} \ha^\dagger_{j,0}\ha^\dagger_{j,1}|vac\> =  \ha_{j,1}^{n} \ha^\dagger_{j,0}\ha^\dagger_{j,1}|vac\> =0. ~~~(n\geq 2)		
	\end{align}
	
	The above identities are translated into our bigraph language as
	\begin{align}\label{path_identity}
		&\begin{tikzpicture}[baseline={([yshift=-.5ex]current bounding box.center)}]
			\node[circle,draw,minimum size=0.7cm] (j) at (0,1.5) {$j$};
			\node[fill, vertex] (v) at (1.8,.8) { };
			\node[fill, vertex] (u) at (1.8,2.2) { };
			\path[line width = 0.8pt, color=red] (j) edge   (u);
			\path[line width = 0.8pt, color=blue] (j) edge  (v);
		\end{tikzpicture}~=~0,~~\nn \\
		&\begin{tikzpicture}[baseline={([yshift=-.5ex]current bounding box.center)}]
			\node[circle,draw,minimum size=0.7cm] (j) at (0,1.5) {$j$};
			\node[fill, vertex] (v) at (1.8,.8) { };
			\node[fill, vertex] (w) at (1.8,2.2) { };
			\draw[thick, dashed] (.7,1.25) to[bend right=50] (.7,1.75);	
			\path[line width = 0.8pt] (j) edge  (v);
			\path[line width = 0.8pt] (j) edge  (w);
			\node at (1.4,1.5) {$n~(\geq 2)$} ;
		\end{tikzpicture}~=~		\begin{tikzpicture}[baseline={([yshift=-.5ex]current bounding box.center)}]
			\node[circle,draw,minimum size=0.7cm] (j) at (0,1.5) {$j$};
			\node[fill, vertex] (v) at (1.8,.8) { };
			\node[fill, vertex] (w) at (1.8,2.2) { };
			\draw[thick, dashed] (.7,1.25) to[bend right=50] (.7,1.75);	
			\path[dotted, line width = 0.8pt] (j) edge  (v);
			\path[dotted, line width = 0.8pt] (j) edge  (w);
			\node at (1.4,1.5) {$n~(\geq 2)$} ;
		\end{tikzpicture}~=~ 0. 
	\end{align} 
	Here, the internal state edge weights $\{|0\>,|1\>, |+\>,|-\> \}$  are denoted as edge colors $\{\textrm{Black, Dotted, Red, Blue}\}$ respectively for the convenience. The amplitude edge weights are omitted. The translation from Eq.~\eqref{qubit_identities_2} to \eqref{path_identity} can be explained more clearly with directed bigraphs (see  Appendix~\ref{comprehensive list}). 
	
	Then, we define effective PM bigraphs (EPM bigraphs) as \emph{bigraphs whose edges always attach to the circles as one of the above forms.}	An example of EPM bigraphs is the $N=2$ bigraph \eqref{N=2_A} with restrictions~\eqref{N=2_amplitudes}, i.e.,
	\begin{align}\label{Bell}
		\begin{tikzpicture}[baseline={([yshift=-.5ex]current bounding box.center)}]
			\node[circle,draw,minimum size=0.6cm] (1) at (0,1.8) {$1$};
			\node[circle,draw,minimum size=0.6cm] (2) at (0,0) {$2$}; 
			\node[fill, vertex] (v) at (3,1.8) { };
			\path[red, line width = 0.8pt] (1) edge (v);
			\path[blue, line width = 0.8pt] (2) edge (v);
			\node[fill, vertex] (u) at (3,0) { };
			\path[blue, line width = 0.8pt ] (1) edge  (u);
			\path[red, line width = 0.8pt] (2) edge (u);
		\end{tikzpicture},
	\end{align}  because all the edges are attached to the circles as the first form of \eqref{path_identity}.  
	
	From the identities~\eqref{path_identity}, we can see a crucial property of EPM bigraphs as follows:
	\begin{tcolorbox}[enhanced jigsaw,colback=black!10!white,boxrule=0pt,arc=0pt]
		\textbf{Property 2.} If a sculpting bigraph is an EPM bigraph, then the final
		state is always fully determined by the PMs of the bigraph.
	\end{tcolorbox}
	The Combination of Properties 1 and 2 provides a convenient strategy to find sculpting operators that generate a specific entangled state. Since we can express an entangled state with a addtion of PMs, \emph{if we can draw an effective PM bigraph which has the same PMs, the bigraph corresponds to a sculpting bigraph that generates the entangled state.} 
	We will show that various qubit $N$-partite genuinely entangled states can be generated with such bigraphs.


	\subsection{Qubit GHZ state}\label{GHZ_qubit_N}

	The sculpting bigraph that generates the $N$-partite GHZ state is given by 
	\begin{align}\label{GHZ_qubit}
		\begin{tikzpicture}[baseline={([yshift=-.5ex]current bounding box.center)}]
			\node[fill, vertex] (1) at (3,5) { } ;
			\node[fill, vertex] (2) at (3,3.5) { };
			\node[fill, vertex] (3) at (3,2) { };
			\draw[dashed, thick] (3,1.5) -- (3,-.5); 
			\node[fill, vertex] (4) at (3,-1) { };    
			\node[circle,draw,minimum size=0.8cm] (X) at (0,5) {1};
			\node[circle,draw,minimum size=0.8cm] (Y) at (0,3.5) {2};
			\node[circle,draw,minimum size=0.8cm] (Z) at (0,2) {3};	
			\draw[dashed, thick] (0,1.3) -- (0,-.3); 	
			\node[circle,draw,minimum size=0.8cm] (W) at (0,-1) {N};	    
			\path[line width = 0.8pt, color=red] (1) edge node {$\frac{1}{\sqrt{2}}$} (X);
			\path[line width = 0.8pt, color=red] (2) edge node {$\frac{1}{\sqrt{2}}$}  (Y);
			\path[line width = 0.8pt, color=red] (3) edge node {$\frac{1}{\sqrt{2}}$}  (Z);
			\path[line width = 0.8pt, color=red] (4) edge node {$\frac{1}{\sqrt{2}}$} (W); 
			\path[line width = 0.8pt, color=blue] (1) edge node {$-\frac{1}{\sqrt{2}}$} (Y);
			\path[line width = 0.8pt, color=blue] (2) edge node {$-\frac{1}{\sqrt{2}}$}  (Z);
			\path[line width = 0.8pt, color=blue] (4) edge node[near start] {$-\frac{1}{\sqrt{2}}$} (X); 
			\draw[dashed, thick] (1.5,1) -- (1.5,0); 
			\draw[line width = 0.8pt,color=blue ] (1.9,1.35) -- (3,2); 	
			\draw[line width = 0.8pt, color=blue] (0.35,-.8) -- (1.2,-.3); 
		\end{tikzpicture},
	\end{align}		
	where the edge weights represent the probability amplitudes and edge colors Red and Blue represent the internal states $|+\>$ and $|-\>$.  This bigraph was also used in Ref.~\cite{chin2021graph} to obtain the GHZ state in LQNs (see bigraph (30) of Ref.~\cite{chin2021graph}).
	
	The sculpting operator $\hat{A}_{N}$ corresponding to \eqref{GHZ_qubit} is
	\begin{align}\label{GHZ_qubit_op}
		\hat{A}_N = \frac{1}{\sqrt{2^N}}&  (\ha_{1,+} - \ha_{2,-})(\ha_{2,+} - \ha_{3,-})\cdots \nn \\
		&\times (\ha_{N-1,+} - \ha_{N,-})(\ha_{N,+} - \ha_{1,-}) \nn \\
		=\frac{1}{\sqrt{2^N}}& \prod_{j=1}^N (\ha_{j,+}-\ha_{j\oplus_N 1,-}),
	\end{align}
	where $\oplus_N$ in the last line is defined as the addition mod $N$. 	
	
	It is simple to verify that the bigraph~\eqref{GHZ_qubit} corresponds to a sculpting operator that generates the GHZ state. First, the edges of~\eqref{GHZ_qubit} attach to circles as the first  of the three graphs in~\eqref{path_identity}. Therefore we see that only the PMs contribute to the final state. Second, the bigraph has two PMs
	\begin{align}\label{GHZ_PM}
		\begin{tikzpicture}[baseline={([yshift=-.5ex]current bounding box.center)}]
			\node[fill, vertex] (1) at (3,2) { } ;
			\node[fill, vertex] (2) at (3,1) { };
			\node[fill, vertex] (3) at (3,0) { };
			\draw[dashed, thick] (0,-0.5) -- (0,-1.5); 
			\node[fill, vertex] (4) at (3,-2) { };    
			\node[circle,draw,minimum size=0.8cm] (X) at (0,2) {$1$};
			\node[circle,draw,minimum size=0.8cm] (Y) at (0,1) {$2$};
			\node[circle,draw,minimum size=0.8cm] (Z) at (0,0) {$3$};	
			\draw[dashed, thick] (2,-0.5) -- (2,-1.5); 	
			\node[circle,draw,minimum size=0.8cm] (W) at (0,-2) {$N$};	    
			\path[line width = 0.8pt,color=red] (1) edge node {$\frac{1}{\sqrt{2}}$} (X);
			\path[line width = 0.8pt,color=red] (2) edge node {$\frac{1}{\sqrt{2}}$} (Y);
			\path[line width = 0.8pt,color=red] (3) edge node {$\frac{1}{\sqrt{2}}$} (Z);
			\path[line width = 0.8pt,color=red] (4) edge node {$\frac{1}{\sqrt{2}}$} (W); 
			\draw[dashed, thick] (1,-1) -- (1,-1.4); 
		\end{tikzpicture}
	\end{align} and
	\begin{align}
		\begin{tikzpicture}[baseline={([yshift=-.5ex]current bounding box.center)}]
			\node[fill, vertex] (1) at (3,4) { } ;
			\node[fill, vertex] (2) at (3,3) { };
			\node[fill, vertex] (3) at (3,2) { };
			\draw[dashed, thick] (3,1.5) -- (3,.5); 
			\node[fill, vertex] (4) at (3,0) { };    
			\node[circle,draw,minimum size=0.8cm] (X) at (0,4) {1};
			\node[circle,draw,minimum size=0.8cm] (Y) at (0,3) {2};
			\node[circle,draw,minimum size=0.8cm] (Z) at (0,2) {3};	
			\draw[dashed, thick] (0,1.3) -- (0,.6); 	
			\node[circle,draw,minimum size=0.8cm] (W) at (0,0) {N}; 
			\path[line width = 0.8pt, color=blue] (1) edge node {$-\frac{1}{\sqrt{2}}$} (Y);
			\path[line width = 0.8pt, color=blue] (2) edge node {$-\frac{1}{\sqrt{2}}$}  (Z);
			\path[line width = 0.8pt, color=blue] (4) edge node[near start] {$-\frac{1}{\sqrt{2}}$} (X); 
			\draw[dashed, thick] (1.5,1.3) -- (1.5,.7); 
			\draw[line width = 0.8pt,color=blue ] (1.9,1.6) -- (3,2); 	
			\draw[line width = 0.8pt, color=blue] (0.35,.2) -- (1.2,.6); 
		\end{tikzpicture}~=~
		\begin{tikzpicture}[baseline={([yshift=-.5ex]current bounding box.center)}]		
			\node[fill, vertex] (1) at (6.5,2) { } ;
			\node[fill, vertex] (2) at (6.5,1) { };
			\node[fill, vertex] (3) at (6.5,0) { };
			\draw[dashed, thick] (3.5,-0.5) -- (3.5,-1.5); 
			\node[fill, vertex] (4) at (6.5,-2) { };    
			\node[circle,draw,minimum size=0.8cm] (X) at (3.5,2) {$1$};
			\node[circle,draw,minimum size=0.8cm] (Y) at (3.5,1) {$2$};
			\node[circle,draw,minimum size=0.8cm] (Z) at (3.5,0) {$3$};	
			\draw[dashed, thick] (6.5,-0.5) -- (6.5,-1.5); 	
			\node[circle,draw,minimum size=0.8cm] (W) at (3.5,-2) {$N$};	    
			\path[line width = 0.8pt,color=blue] (1) edge node {$-\frac{1}{\sqrt{2}}$} (X);
			\path[line width = 0.8pt,color=blue] (2) edge node {$-\frac{1}{\sqrt{2}}$} (Y);
			\path[line width = 0.8pt,color=blue] (3) edge node {$-\frac{1}{\sqrt{2}}$} (Z);
			\path[line width = 0.8pt,color=blue] (4) edge node {$-\frac{1}{\sqrt{2}}$} (W); 
			\draw[dashed, thick] (4.5,-1) -- (4.5,-1.4); 
		\end{tikzpicture}	
	\end{align} (the above equality holds since the dots are identical), which constructs the GHZ state. 
	
	In the operator form, with the identity~\eqref{qubit_identities}, we see that the final state is explicitly given by
	\begin{align}
		\hat{A}_N|Sym_N\> &= \frac{1}{\sqrt{2^N}} \Big(\prod_{j=1}^N\ha_{j,+} + \prod_{j=1}^N\ha_{j,-}\Big)
		|Sym_N\> \nn \\
		&= \frac{1}{\sqrt{2^N}} \Big(\prod_{j=1}^N\ha^\dagger_{j,+} +\prod_{j=1}^N\ha^\dagger_{j,-} \Big)|vac\> \nn \\
		&= \frac{1}{\sqrt{2^{N-1}}}|GHZ_{N,2}\>.
	\end{align} 
	From the normalization factor, we directly see that the success probability becomes $1/2^{N-1}$.

	Note that we can find other sculpting bigraphs for the GHZ state based on \eqref{GHZ_qubit}. While the GHZ state is invariant under the permutation of spatial modes, the bigraph~\eqref{GHZ_qubit} is not. Therefore,  any bigraph with the permuted vertex labels of \eqref{GHZ_qubit} also generates the GHZ state, i.e.,
	\begin{align}\label{GHZ_qubit_perm}
		\begin{tikzpicture}[baseline={([yshift=-.5ex]current bounding box.center)}]
			\node[fill, vertex] (1) at (3,5) { } ;
			\node[fill, vertex] (2) at (3,3.5) { };
			\node[fill, vertex] (3) at (3,2) { };
			\draw[dashed, thick] (3,1.5) -- (3,-.5); 
			\node[fill, vertex] (4) at (3,-1) { };    
			\node[circle,draw,minimum size=0.8cm] (X) at (0,5) {$\s(1)$};
			\node[circle,draw,minimum size=0.8cm] (Y) at (0,3.5) {$\s(2)$};
			\node[circle,draw,minimum size=0.8cm] (Z) at (0,2) {$\s(3)$};	
			\draw[dashed, thick] (0,1.3) -- (0,-.3); 	
			\node[circle,draw,minimum size=0.8cm] (W) at (0,-1) {$\s(N)$};	    
			\path[line width = 0.8pt, color=red] (1) edge node {$\frac{1}{\sqrt{2}}$} (X);
			\path[line width = 0.8pt, color=red] (2) edge node {$\frac{1}{\sqrt{2}}$}  (Y);
			\path[line width = 0.8pt, color=red] (3) edge node {$\frac{1}{\sqrt{2}}$}  (Z);
			\path[line width = 0.8pt, color=red] (4) edge node {$\frac{1}{\sqrt{2}}$} (W); 
			\path[line width = 0.8pt, color=blue] (1) edge node {$-\frac{1}{\sqrt{2}}$} (Y);
			\path[line width = 0.8pt, color=blue] (2) edge node {$-\frac{1}{\sqrt{2}}$}  (Z);
			\path[line width = 0.8pt, color=blue] (4) edge node[near start] {$-\frac{1}{\sqrt{2}}$} (X); 
			\draw[dashed, thick] (1.5,1) -- (1.5,0); 
			\draw[line width = 0.8pt,color=blue ] (1.9,1.35) -- (3,2); 	
			\draw[line width = 0.8pt, color=blue] (0.5,-.7) -- (1.2,-.3); 
		\end{tikzpicture},
	\end{align} 
	under a permutation $\s$ ($\in S_N$). Since the GHZ state is also invariant under the qubit state flip, the exchange of blue and red edges also gives the GHZ state. However, such graph transformations are already included in the above permutation.
	
	It is worth comparing our GHZ solution~\eqref{GHZ_qubit_op} with the solution  given in Ref.~\cite{karczewski2019sculpting},
	\begin{align}\label{GHZ_original}
		\hat{A}_N = \frac{1}{\sqrt{2^N}}\prod_{l=1}^N \Big( \sum_{j=1}^N\ha_{j,0} +\sum_{j=1}^N e^{\frac{2\pi i}{N}(j-l)} \ha_{j,1}\Big).
	\end{align} 
	Most importantly, the consecutive annihilations in \eqref{GHZ_original} do not remove particles from orthogonal modes.  Hence they are very challenging to realize with experimental setups. In contrast, the procedure~\eqref{GHZ_qubit_op} is based on orthogonal modes, so that a single unitary change of basis is sufficient to prepare all the modes from which a single particle is to be removed. On top of that, each  mode in \eqref{GHZ_original} is a weighted superposition of all the initial ones. Understanding the operator from the graph picture,  \eqref{GHZ_original} corresponds to a bigraph with $N^N$ edges. On the other hand, \eqref{GHZ_qubit} corresponds to a bigraph with only $2^N$ edges. Therefore, the scheme described by \eqref{GHZ_qubit} is more effective in the sense that each annihilation operator used there is constructed by superposing just two modes with internal state basis changes.

	\subsection{Qubit W state}\label{W_qubit_N}
	
	A sculpting bigraph for $N$-partite W state can be conceived with one ancillary spatial mode as
	\begin{align}\label{W-state_graph}
		\begin{tikzpicture}[baseline={([yshift=-.5ex]current bounding box.center)}]
			\node[fill, vertex] (1) at (4,5) { } ;
			\node[fill, vertex] (2) at (4,3.5) { };
			\node[fill, vertex] (3) at (4,2) { };
			\draw[dashed, thick] (4,1.5) -- (4,0); 
			\node[fill, vertex] (4) at (4,0) { };   
			\node[fill, vertex] (5) at (4,-1.5) { };   			 
			\node[circle,draw,minimum size=0.8cm] (X) at (0,5) {1};
			\node[circle,draw,minimum size=0.8cm] (Y) at (0,3.5) {2};
			\node[circle,draw,minimum size=0.8cm] (Z) at (0,2) {3};	
			\draw[dashed, thick] (0,1.3) -- (0,.5); 	
			\node[circle,draw,minimum size=0.8cm] (W) at (0,-.0) {$N$};	
			\node[circle,draw,minimum size=0.8cm] (V) at (0,-1.5) {$A$};				    
			\path[line width = 0.8pt, color=red] (1) edge node[above] {$\a$} (X);
			\path[line width = 0.8pt, color=red] (2) edge node[above] {$\a$} (Y);
			\path[line width = 0.8pt, color=red] (3) edge  node[above] {$\a$} (Z);
			\path[line width = 0.8pt, color=red] (4) edge node[above] {$\a$} (W);			
			\path[line width = .8pt] (1) edge node[near start, above] {$\b$}  (V);
			\path[line width = 0.8pt] (2) edge   node[near start,above] {$\b$} (V);
			\path[line width = 0.8pt] (3) edge node[near start,above] {$\b$} (V);
			\path[line width = 0.8pt] (4) edge node[near end,above] {$\b$} (V);			
			\path[line width = 0.8pt, color=blue] (5) edge node[near end] {$\frac{1}{\sqrt{N}}$} (X); 			
			\path[line width = 0.8pt, color=blue] (5) edge node[near end] {$\frac{1}{\sqrt{N}}$}  (Y); 
			\path[line width = 0.8pt, color=blue] (5) edge  node[near end] {$\frac{1}{\sqrt{N}}$} (Z); 	
			\path[line width = 0.8pt, color=blue] (5) edge node[near start] {$\frac{1}{\sqrt{N}}$}  (W); 					
		\end{tikzpicture}.
	\end{align}	
	Edge color Red, Blue, and Black respectively represent the internal state $|+\>,|-\>$ and $|0\>$, and $|\a|^2+|\b|^2=1$. Circle $A$ denotes the ancillary spatial mode. Note that this bigraph shares the same permutation symmetry as the W state, i.e.,  invariance under the permutation of spatial modes. 	
	
	The sculpting operator corresponding to \eqref{W-state_graph} is given by	
	\begin{align}\label{W-state_op}
		&\hat{A}_{N+A}  \nn \\
		&=  \Big(\a\ha_{1+}  +\b \ha_{A0}  \Big) 
		\Big(\a\ha_{2+} +\b \ha_{A0} \Big) \cdots \Big(\a\ha_{N+} +\b \ha_{A0} \Big) \nn  \\
		&~~~~~~~~~~~~\times  \frac{1}{\sqrt{ N}}
		\Big(\ha_{1-} +\ha_{2-}+\cdots +\ha_{N-} \Big) \nn \\
		&= \frac{1}{\sqrt{N}} \Big(\prod_{j=1}^N(\a\ha_{j+} 
		+\b \ha_{A0})\Big)\sum_{k=1}^N\ha_{k-}.
	\end{align}
	The initial state is prepared in a slightly varied way as
	\begin{align}
		|Sym_{N+A}\> \equiv \Big( \prod_{m=1}^N\ha^\dagger_{m0}\ha^\dagger_{m1} \Big)\ha^\dagger_{A0}|vac\>, 
	\end{align} one ancillary boson at the ancillary spatial mode. 
	
	It is as manifest as for the GHZ state case to see the above sculpting bigraph~\eqref{W-state_graph} generates W state. First, the bigraph is an effective bigraph since edges attach to circles as the first and second graphs in \eqref{path_identity}. Second, the above bigraph has $N$ PMs
	\begingroup
	\allowdisplaybreaks
	\begin{align}\label{NW}
		\begin{tikzpicture}[baseline={([yshift=-.5ex]current bounding box.center)}]
			\node[fill, vertex] (1) at (2.5,2) {} ;
			\node[fill, vertex] (2) at (2.5,1) {};
			\node[fill, vertex] (3) at (2.5,0) {};
			\draw[dashed, thick] (0,-0.5) -- (0,-1.5); 
			\node[fill, vertex] (4) at (2.5,-2) {};  
			\node[fill, vertex] (5) at (2.5,-3) {};  
			\node[circle,draw,minimum size=0.8cm] (X) at (0,2) {$1$};
			\node[circle,draw,minimum size=0.8cm] (Y) at (0,1) {$2$};
			\node[circle,draw,minimum size=0.8cm] (Z) at (0,0) {$3$};	
			\draw[dashed, thick] (2.5,-0.5) -- (2.5,-1.5); 	
			\node[circle,draw,minimum size=0.8cm] (W) at (0,-2) {$N$};	  
			\node[circle,draw,minimum size=0.8cm] (K) at (0,-3) {$A$};			
			\path[line width = 0.8pt,color=blue] (1) edge (X);
			\path[line width = 0.8pt,color=red] (2) edge (Y);
			\path[line width = 0.8pt,color=red] (3) edge (Z);
			\path[line width = 0.8pt,color=red] (4) edge (W); 
			\path[line width = 0.8pt] (5) edge (K); 			
			\draw[dashed, thick] (1.25,-1) -- (1.25,-1.4); 
		\end{tikzpicture},\quad 
		& \begin{tikzpicture}[baseline={([yshift=-.5ex]current bounding box.center)}]
			\node[fill, vertex] (1) at (2.5,2) {} ;
			\node[fill, vertex] (2) at (2.5,1) {};
			\node[fill, vertex] (3) at (2.5,0) {};
			\draw[dashed, thick] (0,-0.5) -- (0,-1.5); 
			\node[fill, vertex] (4) at (2.5,-2) {};  
			\node[fill, vertex] (5) at (2.5,-3) {};  
			\node[circle,draw,minimum size=0.8cm] (X) at (0,2) {$1$};
			\node[circle,draw,minimum size=0.8cm] (Y) at (0,1) {$2$};
			\node[circle,draw,minimum size=0.8cm] (Z) at (0,0) {$3$};	
			\draw[dashed, thick] (2.5,-0.5) -- (2.5,-1.5); 	
			\node[circle,draw,minimum size=0.8cm] (W) at (0,-2) {$N$};	  
			\node[circle,draw,minimum size=0.8cm] (K) at (0,-3) {$A$};			
			\path[line width = 0.8pt,color=red] (1) edge (X);
			\path[line width = 0.8pt,color=blue] (2) edge (Y);
			\path[line width = 0.8pt,color=red] (3) edge (Z);
			\path[line width = 0.8pt,color=red] (4) edge (W); 
			\path[line width = 0.8pt] (5) edge (K); 			
			\draw[dashed, thick] (1.25,-1) -- (1.25,-1.4); 
		\end{tikzpicture}, \nn \\  
		\nn \\
		\qquad \cdots\cdots, \qquad \quad  \quad 
		&\begin{tikzpicture}[baseline={([yshift=-.5ex]current bounding box.center)}]
			\node[fill, vertex] (1) at (2.5,2) {} ;
			\node[fill, vertex] (2) at (2.5,1) {};
			\node[fill, vertex] (3) at (2.5,0) {};
			\draw[dashed, thick] (0,-0.5) -- (0,-1.5); 
			\node[fill, vertex] (4) at (2.5,-2) {};  
			\node[fill, vertex] (5) at (2.5,-3) {};  
			\node[circle,draw,minimum size=0.8cm] (X) at (0,2) {$1$};
			\node[circle,draw,minimum size=0.8cm] (Y) at (0,1) {$2$};
			\node[circle,draw,minimum size=0.8cm] (Z) at (0,0) {$3$};	
			\draw[dashed, thick] (2.5,-0.5) -- (2.5,-1.5); 	
			\node[circle,draw,minimum size=0.8cm] (W) at (0,-2) {$N$};	  
			\node[circle,draw,minimum size=0.8cm] (K) at (0,-3) {$A$};			
			\path[line width = 0.8pt,color=red] (1) edge (X);
			\path[line width = 0.8pt,color=red] (2) edge (Y);
			\path[line width = 0.8pt,color=red] (3) edge (Z);
			\path[line width = 0.8pt,color=blue] (4) edge (W); 
			\path[line width = 0.8pt] (5) edge (K); 			
			\draw[dashed, thick] (1.25,-1) -- (1.25,-1.4); 
		\end{tikzpicture},
	\end{align}
	\endgroup	
	which correspond to the W state with an ancillary system.
	
	In the operator form, with the first identity of Eq.~\eqref{qubit_identities} again, we have
	\begin{align}
		&\hat{A}_{N+A} |Sym_{N+A}\> \nn \\
		&= \frac{\a^{N-1}\b}{\sqrt{ N}}( \ha_{1-}\ha_{2+}\cdots \ha_{N+} +\ha_{1+}\ha_{2-}\cdots \ha_{N+} \nn \\
		&\phantom{ffffffff}+\cdots + \ha_{1+}\ha_{2+}\cdots \ha_{N-}) \nn \\
		&\phantom{ffffffff}\times \ha_{A0}\ha^\dagger_{A0}\prod_{m=1}^N\ha^\dagger_{m0}\ha^\dagger_{m1}|vac\> \nn \\
		&= \frac{\a^{N-1}\b}{\sqrt{ N}}( \ha^\dagger_{1-}\ha^\dagger_{2+}\cdots \ha^\dagger_{N+} +\ha^\dagger_{1+}\ha^\dagger_{2-}\cdots \ha^\dagger_{N+} \nn \\
		&\phantom{ffffffff}+\cdots + \ha^\dagger_{1+}\ha^\dagger_{2+}\cdots \ha^\dagger_{N-})|vac\> \nn \\
		&= \a^{N-1}\b|W_N\>.
	\end{align}
	The success probability is $|\a^{N-1}\b|^2$, whose maximal value becomes $\frac{(N-1)^{N-1}}{N^N}$ when $|\a|= \sqrt{\frac{N-1}{N}}$ and $|\b|=\frac{1}{\sqrt{N}}$.
	
	We can also check that this bigraph can be used to generate W state in LQNs with postselection. Indeed, by drawing a bigraph that corresponds to the schemes suggested in Ref.~\cite{bellomo2017n,kim2020efficient}, we obtain the same form of bigraph with \eqref{NW}.
	
	Comparing with the W state generation scheme suggested in Ref.~\cite{karczewski2019sculpting}, we can easily see that our current scheme have accomplished an outstanding improvement. 
	The scheme in Ref.~\cite{karczewski2019sculpting} starts from $4N$ bosons in $2N$ modes and goes through two steps of sculpting to generate the final $N$-partite W state. 
	On the other hand, using the graph mapping technique, we have obtained a much more efficient $N$-partite W-state generation scheme just with $2N+1$ bosons in $N+1$ spatial modes and one simple step of sculpting. 	
	
	\subsection{$N=3$ Type 5 states}\label{Type_5}
	
	The genuinely entangled states that we have discussed so far have some convenient symmetries, which admit relatively simple sculpting bigraphs for generating them. However, we can also conceive less symmetric entangled states with the support of ancillary modes. It is always achieved by any EPM bigraph that connects all the dots to \encircle{$A_j$} with black or dotted edges.
	
	As an example, we present a EPM bigraph for a tripartite system that generates
	a superposition of the $N=3$ GHZ and W states, which is called the $N=3$ Type 5 state in Ref.~\cite{blasiak2022arbitrary}. The EPM bigraph used here includes three ancillae to construct such a sculpting operator: 
	
	\begin{align}\label{3state_graph}
		\begin{tikzpicture}[baseline={([yshift=-.5ex]current bounding box.center)}]
			\node[fill, vertex] (1) at (4,5) { } ;
			\node[fill, vertex] (2) at (4,3.5) { };
			\node[fill, vertex] (3) at (4,2) { };
			\node[fill, vertex] (4) at (4,.5) { }; 
			\node[fill, vertex] (5) at (4,-1) { };   
			\node[fill, vertex] (6) at (4,-2.5) { };   			 
			\node[circle,draw,minimum size=0.8cm] (X) at (0,5) {1};
			\node[circle,draw,minimum size=0.8cm] (Y) at (0,3.5) {2};
			\node[circle,draw,minimum size=0.8cm] (Z) at (0,2) {3};	
			\node[circle,draw,minimum size=0.8cm] (W) at (0,.5) {$A$};		
			\node[circle,draw,minimum size=0.8cm] (V) at (0,-1) {$B$};	
			\node[circle,draw,minimum size=0.8cm] (U) at (0,-2.5) {$C$};				    
			\path[line width = 0.8pt, color=red] (1) edge  node {$\frac{1}{\sqrt{2}}$}  (X);
			\path[line width = 0.8pt, color=red] (2) edge  node {$\frac{1}{\sqrt{2}}$} (Y);
			\path[line width = 0.8pt, color=red] (3) edge  node {$\frac{1}{\sqrt{2}}$} (Z);
			\path[line width = .8pt] (5) edge  node {$\frac{1}{\sqrt{3}}$}  (V);
			\path[line width = 0.8pt] (6) edge  node[near end] {$\frac{1}{\sqrt{3}}$} (U);
			\path[line width = 0.8pt] (1)  edge  node[near start] {$\frac{1}{\sqrt{2}}$}  (W);
			\path[line width = 0.8pt] (2) edge  node[near start] {$\frac{1}{\sqrt{2}}$} (V);
			\path[line width = 0.8pt] (3) edge  node[near start] {$\frac{1}{\sqrt{2}}$} (U); 			
			\path[line width = 0.8pt, color=blue] (4) edge node[near end] {$-\frac{1}{\sqrt{2}}$} (X); 
			\path[line width = 0.8pt, color=blue] (5) edge node {$-\frac{1}{\sqrt{3}}$} (Y);
			\path[line width = 0.8pt, color=blue] (6) edge node[near start] {$-\frac{1}{\sqrt{3}}$} (Z);						\path[line width = 0.8pt, color=black] (4) edge node[near start] {$\frac{1}{\sqrt{3}}$}  (U);			
			\path[line width = .8pt] (5) edge   node[near start] {$\frac{1}{\sqrt{3}}$}  (W);
			\path[line width = 0.8pt] (6) edge  node[near start] {$\frac{1}{\sqrt{3}}$} (V);	
		\end{tikzpicture}.
	\end{align}	
	In the above bigraph, the amplitude weights are omitted under the assumption that they are nonzero and satisfy the normalization conditions. 
	
	The corresponding sculpting operator of the above bigraph is given by 
	\begin{align}\label{arbit}
		&\hat{A}_{3+A,B,C} \nn \\
		&= \frac{(\ha_{1+} +\ha_{A0})}{\sqrt{2}}\frac{(\ha_{2+} +\ha_{B0})}{\sqrt{2}} \frac{(\ha_{3+} +\ha_{C0})}{\sqrt{2}}\nn \\
		&~~\times\frac{(\ha_{C0} - \ha_{1-})}{\sqrt{2}}\frac{(\ha_{A0} +\ha_{B0} - \ha_{2-})}{\sqrt{3}}\frac{(\ha_{B0}+\ha_{C0} - \ha_{3-})}{\sqrt{3}}.
	\end{align}

	When the above sculpting operator is applied to the initial state 
	\begin{align}
		|Sym_{3+A,B,C} \equiv \ha^\dagger_{A0}\ha^\dagger_{B0}\ha^\dagger_{C0}\prod_{m=1}^3\ha^\dagger_{m0}\ha^\dagger_{m1}|vac\>,     
	\end{align}
	we have 
	\begin{align}\label{N=3_arbit_final}
		&\hat{A}_{3+A,B,C}|Sym_{3+A,B,C}\> \nn \\
		&= \frac{1}{12}(\ha_{1+}\ha_{2+}\ha_{3+} 
		- \ha_{1-}\ha_{2+}\ha_{3+} +\ha_{1-}\ha_{2+}\ha_{3-} \nn \\
		&~~~
		+\ha_{1-}\ha_{2-}\ha_{3+} -\ha_{1-}\ha_{2-}\ha_{3-})   \prod_{m=1}^3\ha^\dagger_{m0}\ha^\dagger_{m1}|vac\> \nn \\
		&= \frac{1}{12}(\ha^\dagger_{1+}\ha^\dagger_{2+}\ha^\dagger_{3+} 
		+ \ha^\dagger_{1-}\ha^\dagger_{2+}\ha^\dagger_{3+} +\ha^\dagger_{1-}\ha^\dagger_{2+}\ha^\dagger_{3-} \nn \\
		&~~~
		+\ha^\dagger_{1-}\ha^\dagger_{2-}\ha^\dagger_{3+} +\ha^\dagger_{1-}\ha^\dagger_{2-}\ha^\dagger_{3-})|vac\> \nn \\
		&=\frac{1}{12}(|+++\> + |-++\> + |-+-\> + |--+\> + |---\>).   
	\end{align}
	See that the five states with nonzero amplitudes in the final line of the above equation constitute the set of bases that can transform to any tripartite state under local operations~\cite{acin2000generalized}. The final state we just obtained is categorized as $N=3$ Type 5 state in Ref.~\cite{aaronson2011computational}.  
	
	This $N=3$ example shows that the graph method has potential to design other general form of multipartite entangled states with heralding detectors.
	
	\section{Qudit entanglement: GHZ state}\label{qudit}
	
	Our graph picture also provides a useful insight to find sculpting operators for the general $qudit$ systems. We will present in this section a sculpting bigraph for the qudit $N$-partite GHZ state, which has a generalized form of the qubit GHZ bigraph~\eqref{GHZ_d_graph}.

\begin{figure}
		\centering
		\includegraphics[width=7.5cm]{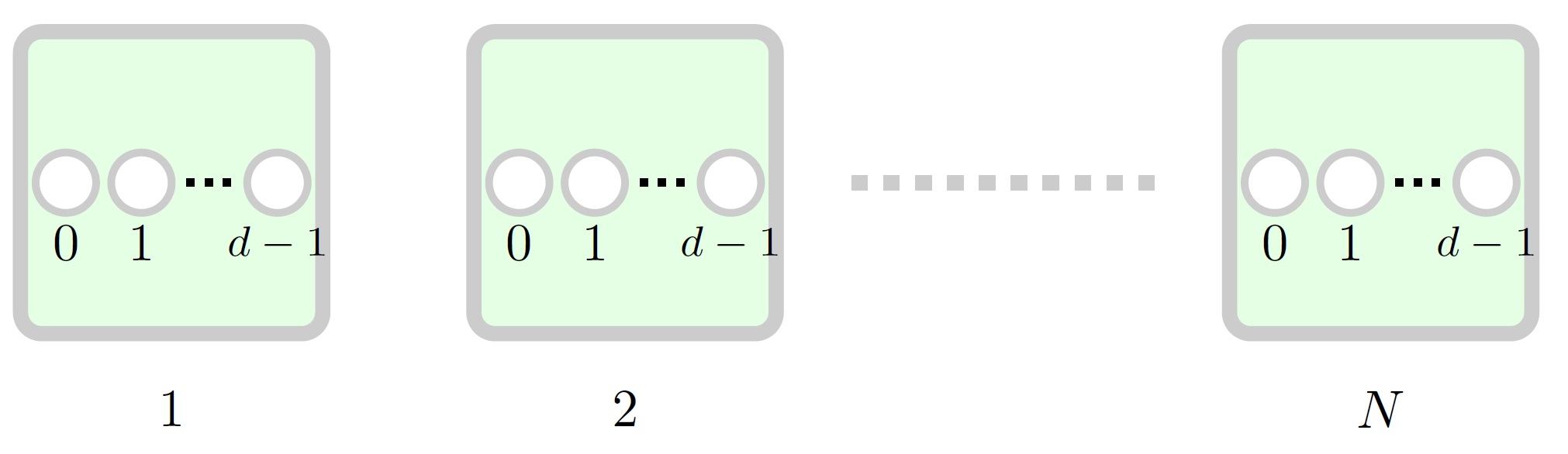}
		\caption{The initial state  $|Sym_{N.d}\>$ of $dN$ bosons in $N$ spatial modes. Each spatial mode has $d$ bosons, which have  mutually orthogonal internal states $|0\>, |1\>,\cdots, |d-1\> $. } 
		\label{fig_initial_qudit}
	\end{figure}

	The qudit state is represented by a $d$-dimensional internal degree of freedom $s$ $(\in \{0,1,\cdots, d\})$ of bosons. To construct $N$ partite qudit genuinely entangled states, we initially distribute $dN$ bosons into $N$ spatial modes so that exactly $d$ bosons with mutually orthogonal internal states belong to a spatial mode (see Fig.~\ref{fig_initial_qudit}). Hence, the initial state is given by 
	\begin{align}\label{initial}
		|Sym_{N,d}\> \equiv \prod_{j=1}^N(\ha_{j,0}\ha_{j,1}\cdots \ha_{j,d})|vac\>.
	\end{align} 
	Here $|Sym_{N,d}\>$ denotes the $N$ spatial mode maximally symmetric state with a $d$-dimensional internal degree of freedom.	
	
	The sculpting operator
	\begin{align}
		\hat{A}_N= \prod_{l=1}^{(d-1)N}\hat{A}^{(l)}
	\end{align}
	must be set to extract $(d-1)$ bosons per spatial mode so that one boson per spatial mode in the final state determines the qudit state of each subsystem.
	All in all, the sculpting protocol is modified for qudits as follows:
	\begin{tcolorbox}[enhanced jigsaw,colback=black!10!white,boxrule=0pt,arc=0pt]	
		\textbf{Sculpting protocol of qudits}
		\begin{enumerate}
			\item Initial state: We prepare the maximally symmetric state $|Sym_{N,d}\>$ of $dN$ bosons, i.e., each boson has different states (either spatial or internal) with each other as Eq.~\eqref{initial}. See Fig.~\ref{fig_initial_qudit}. 
			\item Operation: We apply the sculpting  operator $\hat{A}_N$
			to the initial state $|Sym_{N,d}\>$. The sculpting operator must be set to extract $(d-1)$ bosons per spatial mode. 
			\item Final state: The final state can be fully separable, partially separable, or genuinely entangled.
		\end{enumerate}
	\end{tcolorbox}

	Now we provide a sculpting operator that generates the $N$-partite GHZ state of $d$-level systems, denoted as $|GHZ_{N,d}\>$, by generalizing the qubit GHZ sculpting operator in Sec.~\ref{GHZ_qubit_N}.
	
	First, by generalizing the $d=2$ basis set $\{|+\>, |-\>\}$ for the internal states of the sculpting operators, we choose the arbitrary $d$-dimensional basis set  $\{|\tilde{k}\>\}_{k=0}^{d-1}$ where
	\begin{align}\label{basis_FT}
		|\tilde{k}\> = \frac{1}{\sqrt{d}} \Big(|0\> + \omega^k|1\>+\omega^{2k}|2\> + \cdots +\omega^{(d-1)k}|d-1\> \Big) 
	\end{align} 
	$(\omega = e^{i\frac{2\pi k}{d}})$ for the internal states of the sculpting operators. 
	
	Second, we use an overlap of $(d-1)$ copies of the graph~\eqref{GHZ_qubit} for the sculpting bigraph,  i.e., 
	the following bigraph corresponds to the sculpting operator for the GHZ state:    
	\begin{align}\label{GHZ_d_graph}
		\begin{tikzpicture}[baseline={([yshift=-.5ex]current bounding box.center)}]
			\node[circle,draw,minimum size=0.5cm, fill=gray] (1) at (3,5) { } ;
			\node[circle,draw,minimum size=0.5cm,fill=gray] (2) at (3,3.5) { };
			\node[circle,draw,minimum size=0.5cm,fill=gray] (3) at (3,2) { };
			\draw[dashed, thick] (3,1.5) -- (3,-.5); 
			\node[circle,draw,minimum size=0.5cm,fill=gray] (4) at (3,-1) { };    
			\node[circle,draw,minimum size=0.8cm] (X) at (0,5) {1};
			\node[circle,draw,minimum size=0.8cm] (Y) at (0,3.5) {2};
			\node[circle,draw,minimum size=0.8cm] (Z) at (0,2) {3};	
			\draw[dashed, thick] (0,1.3) -- (0,-.3); 	
			\node[circle,draw,minimum size=0.8cm] (W) at (0,-1) {N};	    
			\path[line width = 0.8pt, color=red] (1) edge node {$\frac{1}{\sqrt{2}}$}  (X);
			\path[line width = 0.8pt, color=red] (2) edge node {$\frac{1}{\sqrt{2}}$}  (Y);
			\path[line width = 0.8pt, color=red] (3) edge node {$\frac{1}{\sqrt{2}}$}  (Z);
			\path[line width = 0.8pt, color=red] (4) edge node {$\frac{1}{\sqrt{2}}$} (W); 
			\path[line width = 0.8pt, color=blue] (1) edge node {$-\frac{1}{\sqrt{2}}$} (Y);
			\path[line width = 0.8pt, color=blue] (2) edge node {$-\frac{1}{\sqrt{2}}$}  (Z);
			\path[line width = 0.8pt, color=blue] (4) edge node[near start] {$-\frac{1}{\sqrt{2}}$} (X); 
			\draw[dashed, thick] (1.5,1) -- (1.5,0); 
			\draw[line width = 0.8pt,color=blue ] (1.9,1.35) -- (2.8,1.88); 	
			\draw[line width = 0.8pt, color=blue] (0.35,-.8) -- (1.2,-.3); 
		\end{tikzpicture} \nn \\
		(\textrm{red: $|\tu\>$, blue: $|\widetilde{d-1}\>$})
	\end{align}	where a gray circle represents a group of $(d-1)$ identical vertices that have the same edges.
	For example, when $N=3$ and $d=4$, the above graph is explicitly drawn as
	\begin{align}\label{n=3_qutrit}
		\begin{tikzpicture}[baseline={([yshift=-.5ex]current bounding box.center)}]
			\node[fill, vertex] (1) at (3,3.3) { } ;
			\node[fill, vertex] (2) at (3,1.8) { };
			\node[fill, vertex] (3) at (3,0.3) { };
			\node[fill, vertex] (1') at (3,2.7) { } ;
			\node[fill, vertex] (2') at (3,1.2) { };
			\node[fill, vertex] (3') at (3,-.3) { };
			\node[fill, vertex] (1'') at (3,3) { } ;
			\node[fill, vertex] (2'') at (3,1.5) { };
			\node[fill, vertex] (3'') at (3,0) { };
			\node[circle,draw,minimum size=0.8cm] (X) at (0,3) {1};
			\node[circle,draw,minimum size=0.8cm] (Y) at (0,1.5) {2};
			\node[circle,draw,minimum size=0.8cm] (Z) at (0,0) {3};	
			\path[line width = 0.8pt, color=red] (1) edge  (X);
			\path[line width = 0.8pt, color=red] (2) edge  (Y);
			\path[line width = 0.8pt, color=red] (3) edge   (Z);
			\path[line width = 0.8pt, color=blue] (1) edge    (Y);
			\path[line width = 0.8pt, color=blue] (2) edge   (Z);
			\path[line width = 0.8pt, color=blue] (3) edge  (X);
			\path[line width = 0.8pt, color=red] (1') edge  (X);
			\path[line width = 0.8pt, color=red] (2') edge  (Y);
			\path[line width = 0.8pt, color=red] (3') edge   (Z);
			\path[line width = 0.8pt, color=blue] (1') edge    (Y);
			\path[line width = 0.8pt, color=blue] (2') edge   (Z);
			\path[line width = 0.8pt, color=blue] (3') edge  (X);
			\path[line width = 0.8pt, color=red] (1'') edge  (X);
			\path[line width = 0.8pt, color=red] (2'') edge  (Y);
			\path[line width = 0.8pt, color=red] (3'') edge   (Z);
			\path[line width = 0.8pt, color=blue] (1'') edge    (Y);
			\path[line width = 0.8pt, color=blue] (2'') edge   (Z);
			\path[line width = 0.8pt, color=blue] (3'') edge  (X);	
			\node[circle, draw=gray, minimum size=.85cm] (a) at (3,3) { } ;
			\node[circle, draw=gray, minimum size=.85cm] (a) at (3,1.5) { } ;
			\node[circle, draw=gray, minimum size=.85cm] (a) at (3,0) { } ;		
		\end{tikzpicture}.
	\end{align}	
	
	The sculpting operator that corresponds to the bigraph~\eqref{GHZ_d_graph} is given by
	\begin{align}\label{GHZ_d}
		&\hat{A}_{N,d} \nn \\
		&=\Big(\frac{1}{\sqrt{2}}\Big)^{(d-1)N}(\ha_{1,\tu}-\ha_{2,\widetilde{d-1}})^{d-1} (\ha_{2,\tu}-\ha_{3,\widetilde{d-1}})^{d-1} \nn \\
		&\phantom{dfdfdf}\times (\ha_{3,\tu}-\ha_{4,\widetilde{d-1}})^{d-1} \cdots\times  (\ha_{N,\tu}-\ha_{1,\widetilde{d-1}})^{d-1}.
	\end{align}	
	We verify that the above operator constructs the qudit $N$-partite GHZ state in Appendix~\ref{qudit_GHZ}. To catch the sense of how the graph~\eqref{GHZ_d_graph} works, we explictly explain the qutrit case ($d=3$) here.

	\subsection{Qutrit GHZ state}
	For a qutrit system, the  basis set~\eqref{basis_FT} is given by
	\begin{align}
		\{&|\tu\> = \frac{1}{\sqrt{3}}(|0\>+|1\>+|2\>) , \nn \\
		&|\tilde{1}\> = \frac{1}{\sqrt{3}}(|0\>+e^{i\frac{2\pi}{3}}|1\>+e^{i\frac{4\pi}{3}}|2\>), \nn \\
		&|\tilde{2}\>=   \frac{1}{\sqrt{3}}(|0\>+e^{i\frac{4\pi}{3}}|1\>+e^{i\frac{2\pi}{3}}|2\>) \}. 
	\end{align} 
	Then, we directly check that the following identities hold:
	\begin{align}\label{identity_qutrit}
		(\ha_{j,\tu})^2\ha^\dagger_{j,0}\ha^\dagger_{j,1}\ha^\dagger_{j,2} &= \frac{2}{\sqrt{3}}\ha^\dagger_{j,\tu}, \nn \\
		\ha_{j,\tu}\ha_{j,\td}\ha^\dagger_{j,0}\ha^\dagger_{j,1}\ha^\dagger_{j,2}  &= - \frac{1}{\sqrt{3}} \ha^\dagger_{j,\t2}, \nn \\
		(\ha_{j,\td})^2\ha^\dagger_{j,0}\ha^\dagger_{j,1}\ha^\dagger_{j,2} &= \frac{2}{\sqrt{3}}\ha^\dagger_{j,\td}.  
	\end{align}
	(note that the above identities are also obtained from Eq.~\eqref{identity_qudit} with $d=3$).
	From the second identity of the above, we can see that
	\begin{align}\label{identity_qutrit_corol}
		(\ha_{j,\tu})^2\ha_{j,\td}\ha^\dagger_{j,0}\ha^\dagger_{j,1}\ha^\dagger_{j,2}  = 		\ha_{j,\tu}(\ha_{j,\td})^2\ha^\dagger_{j,0}\ha^\dagger_{j,1}\ha^\dagger_{j,2}  = 0
	\end{align} also holds.
	The graph~\eqref{GHZ_d_graph} for $d=3$ is now drawn as
	\begin{align} 
		\begin{tikzpicture}[baseline={([yshift=-.5ex]current bounding box.center)}]
			\node[circle, draw=gray, minimum size=.8cm] (a) at (3,5) { } ;
			\node[circle, draw=gray, minimum size=.8cm] (a) at (3,3.5) { } ;
			\node[circle, draw=gray, minimum size=.8cm] (a) at (3,2) { } ;
			\node[circle, draw=gray, minimum size=.8cm] (a) at (3,-1) { } ;	
			\node[fill, vertex] (1) at (3,5.2) { } ;
			\node[fill, vertex] (1') at (3,4.8) { } ;
			\node[fill, vertex] (2) at (3,3.7) { };
			\node[fill, vertex] (2') at (3,3.3) { };
			\node[fill, vertex] (3) at (3,2.2) { };
			\node[fill, vertex] (3') at (3,1.8) { };
			\draw[dashed, thick] (3,1.5) -- (3,-.5); 
			\node[fill, vertex] (4) at (3,-.8) { };
			\node[fill, vertex] (4') at (3,-1.2) { }; 
			\node[circle,draw,minimum size=0.8cm] (X) at (0,5.1) {1};
			\node[circle,draw,minimum size=0.8cm] (Y) at (0,3.6) {2};
			\node[circle,draw,minimum size=0.8cm] (Z) at (0,2.1) {3};	
			\draw[dashed, thick] (0,1.3) -- (0,-.3); 	
			\node[circle,draw,minimum size=0.8cm] (W) at (0,-1) {N};	    
			\path[line width = 0.8pt, color=red] (1) edge  (X);
			\path[line width = 0.8pt, color=red] (1') edge  (X);
			\path[line width = 0.8pt, color=red] (2) edge   (Y);
			\path[line width = 0.8pt, color=red] (2') edge   (Y);
			\path[line width = 0.8pt, color=red] (3) edge   (Z);
			\path[line width = 0.8pt, color=red] (3') edge   (Z);
			\path[line width = 0.8pt, color=red] (4) edge (W); 
			\path[line width = 0.8pt, color=red] (4') edge (W); 
			\path[line width = 0.8pt, color=blue] (1) edge  (Y);
			\path[line width = 0.8pt, color=blue] (1') edge  (Y);
			\path[line width = 0.8pt, color=blue] (2) edge   (Z);
			\path[line width = 0.8pt, color=blue] (2') edge   (Z);
			\path[line width = 0.8pt, color=blue] (4) edge  (X); 
			\path[line width = 0.8pt, color=blue] (4') edge  (X); 
			\draw[dashed, thick] (1.5,1) -- (1.5,0); 
			\draw[line width = 0.8pt,color=blue ] (1.7,1.4) -- (3,2.2); 	
			\draw[line width = 0.8pt,color=blue ] (1.7,1.3) -- (3,1.8); 	
			\draw[line width = 0.8pt, color=blue] (0.37,-.8) -- (1.3,-.4); 
			\draw[line width = 0.8pt, color=blue] (0.33,-.75) -- (1.3,-.2); 
		\end{tikzpicture} \nn \\
		\textrm{(red: $\tu$, blue: $\td$)}
	\end{align}
	and the corresponding sculpting operator becomes
	\begin{align}
		&\hat{A}_{N,3}\nn \\
		&= \Big(\frac{1}{\sqrt{2}}\Big)^{2N}(\ha_{1,\tu}-\ha_{2,\widetilde{2}})^{2} (\ha_{2,\tu}-\ha_{3,\widetilde{2}})^{2} \nn \\
		&\phantom{fdfdfdfdfd}\times (\ha_{3,\tu}-\ha_{4,\widetilde{2}})^{2}\times  \cdots\times  (\ha_{N,\tu}-\ha_{1,\widetilde{2}})^{2}.  
	\end{align}
	Then, the final state is given by 
	\begingroup
	\allowdisplaybreaks
	\begin{align}
		&\hat{A}_{N,3}|Sym_{N,3}\> \nn \\
		&= \Big(\frac{1}{\sqrt{2}}\Big)^{2N}\Big(
		\prod_{r=1}^N(\ha_{r,\tu})^2 +\prod_{s=1}^N(-2\ha_{s\tu}\ha_{s\td})+ \prod_{t=1}^N(\ha_{t,\td})^2\Big) \nn \\
		&~~~\times \prod_{p=1}^N \ha_{p,0}^\dagger\ha_{p,1}^\dagger\ha_{p,2}^\dagger |vac\> \nn \\
		&= \Big(\frac{1}{\sqrt{3}}\Big)^{N}\Big(  \prod_{r=1}^N\ha_{r,\tu}^\dagger  +\prod_{s=1}^N\ha_{s\tilde{1}}^\dagger + \prod_{t=1}^N\ha_{t,\td}^\dagger  \Big) |vac\> \nn \\
		&= \Big(\frac{1}{\sqrt{3}}\Big)^{N-1}|GHZ_{N,3}\> . 
	\end{align}
	\endgroup
	The second line is obtained by Eq.~\eqref{identity_qutrit_corol} and the third by Eq.~\eqref{identity_qutrit}. The success probability is $1/3^{N-1}$. 
	
	\paragraph*{Remark.---} To the best of our knowledge, our scheme needs much less bosons than any other heralded schemes for the qudit GHZ state. For example, the scheme in Ref.~\cite{paesani2021scheme} that needs $(2d+1)$ particles for the $d$-level bipartite GHZ state (i.e., the Bell state) and 25 particles for the 3-level tripartite GHZ state in their optimized method. Our sculpting bigraphs just need $2d$ and 9 particles respectively.
	
	$ $\\
	
	\section{Heralded scheme of Sculpting protocols in linear optics: Bell state example}\label{experiments}
	
	In Sec.~\ref{qubit_graphs} and \ref{qudit}, we have presented sculpting bigraphs that generate genuinely multipartite entangled states. 
		Therefore, if we know how to build spatially overlapped subtraction operators by heralding, we can directly design heralded entanglement generation circuits by combining these operators as the structure of sculpting bigraphs.
	
		There are general schemes in optics~\cite{kim2008scheme,zavatta2009experimental,parigi2007probing,ourjoumtsev2006generating}  to establish subtraction operators of bosons. Based on such methods, Ref.~\cite{karczewski2019sculpting} proposed an optical scheme for constructing sculpting operators. More recently, Ref.~\cite{zaw2022sculpting} suggested an experimental scheme with arithmetic subtractions of trapped ions (which are near-deterministic operations established in Ref.~\cite{um2016phonon} that work unitarily except when the system is in the vacuum) to generate the GHZ state with the sculpting operator~\eqref{GHZ_original} in  Ref.~\cite{karczewski2019sculpting}.  Ref.~\cite{karczewski2019sculpting} and \cite{zaw2022sculpting} used dual-rail encoding and binomial encoding respectively.
	
		Here, we briefly explain an alternative linear optical circuit designed to implement spatially overlapped subtraction operators by heralding, which becomes building blocks of heralded schemes that generates entanglement. Then, as a proof of concept, we design a heralded Bell state generation scheme by  utilizing the heralded subtraction operator and the Bell state bigraph~\eqref{Bell}. 
		The optical elements used in the circuit can be applied to more general multipartite entagled states. For a more comprehensive explanation and solutions on the construction of heralded schemes from our sculpting schemes, see Ref.~\cite{chin2023from}. 
	
		Since all the qubit sculpting schemes in our work correspond to EPM bigraphs whose final states are determined by the identities~\eqref{qubit_identities} and \eqref{qubit_identities_2}, we need to find heralded optical circuits that corresponds to those identities. The basic elements of our optical schemes are
		polarizing beam splitters (PBSs) and half-wave plates (HWPs), hence highly feasible.
		It also implies that our translation rule can be applied to any bosonic systems with operators that play the roles of PBSs and HWPs.

		In our setup, we encode the internal boson states $\{|0\>, |1\>, |+\>,|-\> \}$ as the polarization of photons $\{|D\>, |A\>, |H\>,|V\> \}$ where ($D$=diagonal, $A$=antidiagonal, $H$= horizotal, $V$=vertical). The polarized states have the following relations:
		\begin{align}
			&\<D|A\> = \<H|V\>=0, \nn \\
			&|H\> = \frac{1}{\sqrt{2}}(|D\>+|A\>),~~~|V\> = \frac{1}{\sqrt{2}}(|D\> -|A\>).
		\end{align}
		Then the initial state is given by
		\begin{align}
			|Sym_{N}\> = \prod_{j=1}^N \ha^\dagger_{j,D}\ha^\dagger_{j,A}|vac\>.
		\end{align}

		We first consider the heralded optical circuit for the identity~\eqref{qubit_identities}, which can be rewritten as
		\begin{align}
			\ha_{\pm}\frac{(\ha^{\dagger 2}_+ - \ha^{\dagger 2}_-)}{2}|vac\>  = \pm \ha^\dagger_{\pm}|vac\>.  
		\end{align} 
		We can perform the above operation with the following optical circuit:
		\begin{align}\label{sub_opt}
			\includegraphics[width=8cm]{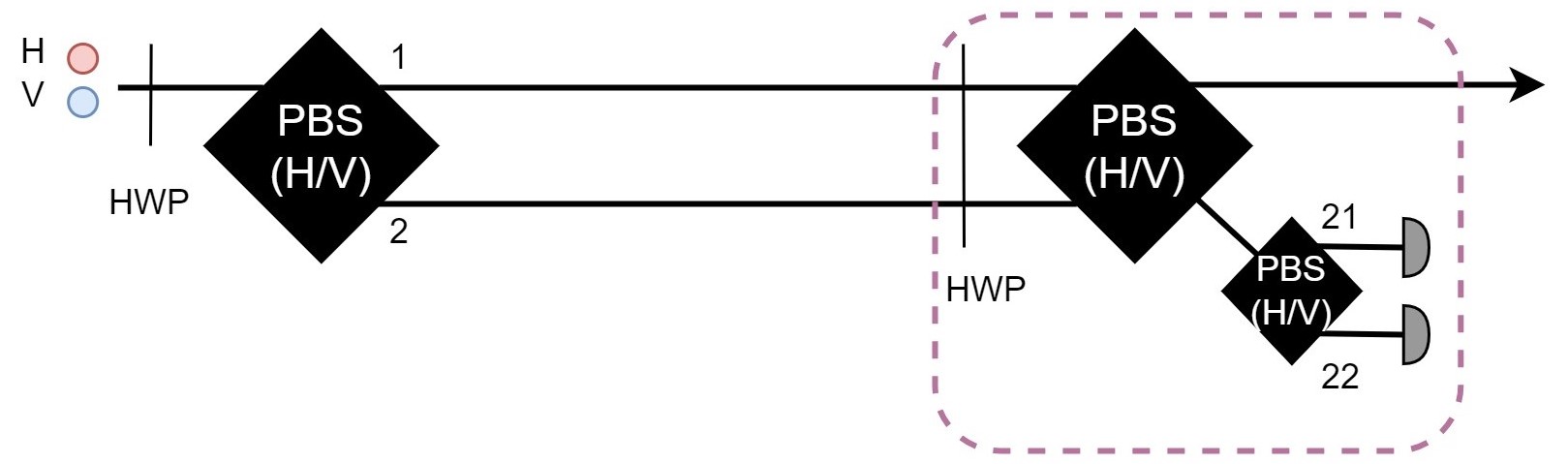}
		\end{align} where the one-photon detection at $21$ and $22$  correspond to $\ha_{+}$ and $\ha_{-}$ respectively. 
		In the above circuit, PBSs transform photons as
		\begin{align}
			\includegraphics[width=.4\textwidth]{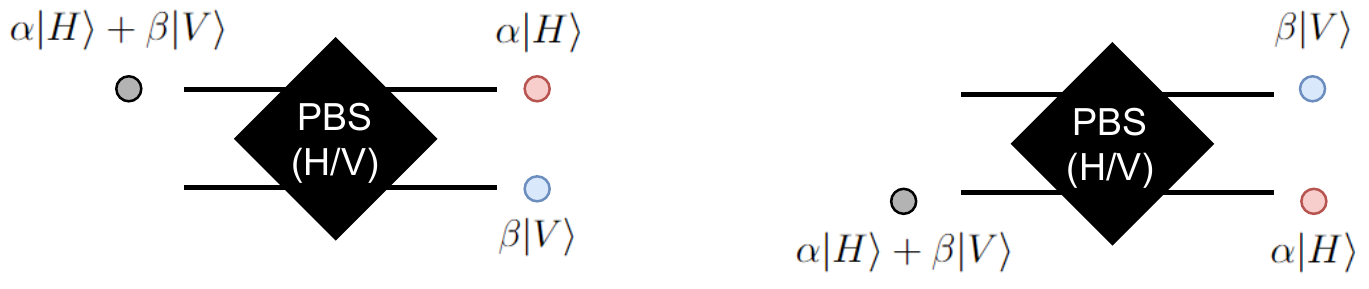}, \nn 
		\end{align}
		and HWPs as  $\{H,V\} \leftrightarrow \{D,A\}$.
	
		\textbf{Step-by-step explanation}
		\begin{enumerate}
			\item The first HWP rotates the photon state basis:
			\begin{align}
				\ha^\dagger_{H}\ha^\dagger_V \to \ha^\dagger_{D}\ha^\dagger_A = \frac{1}{2}(\ha^{\dagger 2}_H - \ha^{\dagger 2}_V)
			\end{align}
			\item The first PBS divides the photon paths according to the internal states:
			\begin{align}
				\frac{1}{2}(\ha^{\dagger 2}_H - \ha^{\dagger 2}_V) \to \frac{1}{2}(\ha^{\dagger 2}_{1,H} - \ha^{\dagger 2}_{2,V}) 
			\end{align}
			where 1 and 2 denote the upper and lower paths of the PBS.
			\item The dashed purple box subtracts $\ha^\dagger_{1,H}$ or $\ha^\dagger_{2,V}$ by heralding in the last PBS and send the remained one to the initial mode:
			\begin{align}
				&\frac{1}{2}(\ha^{\dagger 2}_{1,H} - \ha^{\dagger 2}_{2,V})\nn \\
				&\xrightarrow{HWP}\frac{1}{2}(\ha^{\dagger 2}_{1,D} - \ha^{\dagger 2}_{2,A}) \nn \\
				&\xrightarrow{PBS}\frac{1}{4}\Big( (\ha^\dagger_{1,H}+\ha^\dagger_{2,V})^2 - (\ha^\dagger_{2,H}-\ha^\dagger_{1,V})^2\Big) \nn \\
				&\xrightarrow{P.S.} \ha^\dagger_{1,H}\ha^\dagger_{22,V}~or~ \ha^\dagger_{1,V}\ha^\dagger_{21,H}
			\end{align}
			In the last PBS, the detection of a photon with $V$ in the upper mode 21 ($H$ in the lower mode 22) heralds the final state $\ha^\dagger_{H}$ ($\ha^\dagger_{V}$). 
		\end{enumerate}
		In the above process, the dashed purple box plays the role of the heralded subtraction operator. 
	
		We can deform the subtractor to design a heralded optical circuit for spatially overlapped subtraction operators such as $(\ha_{1,+} -\ha_{2,-})$. Instead of attaching both wires of the subtractor to the same mode as in~\eqref{sub_opt}, we now attach two wires to different modes 1 and 2 as in
		\begin{align}\label{subtraction_opt}
			\includegraphics[width=6cm]{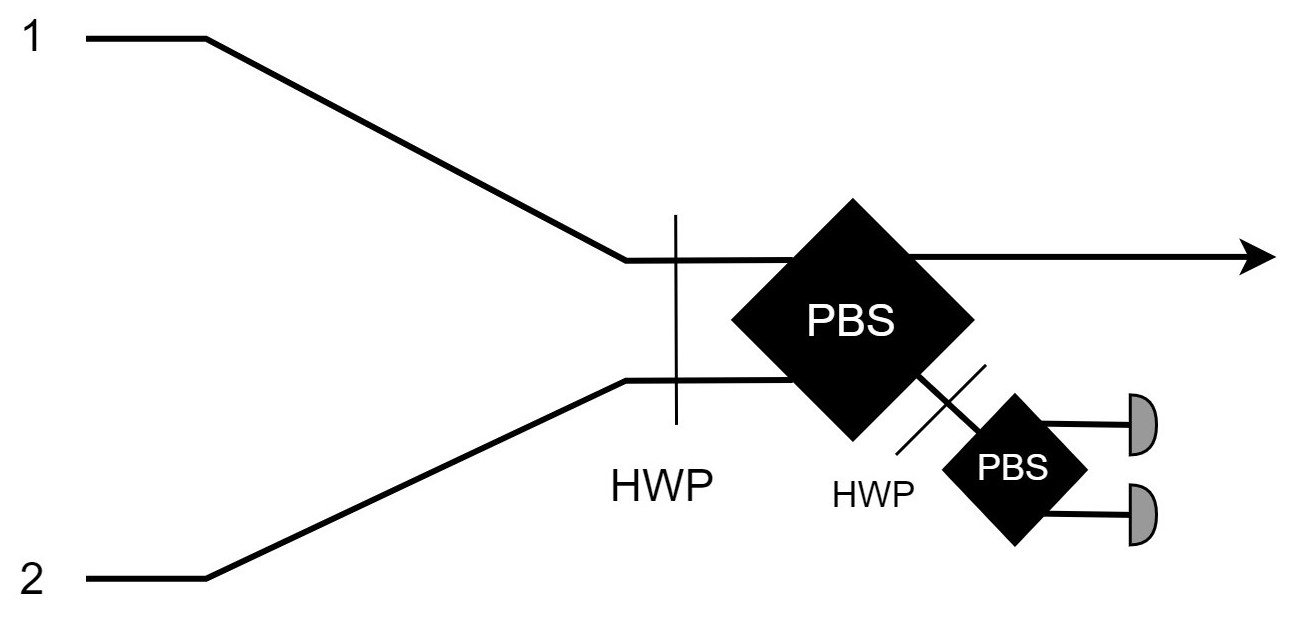}
		\end{align}
		so that it plays the role of a spatially subtraction operator. A crucial difference of the subtraction operator in the above from that in \eqref{sub_opt} is the need of an HWP between two PBSs, which makes possible the subtraction of different internal states from difference input modes. 
	
		As a proof of concept that the above heralded operator plays the role of spatially overlapped subtraction operator, we provide a Bell-state generation scheme by employing the heralded subtraction operator based on the sculpting bigraph~\eqref{Bell},
		\begin{align}\label{bell_bigraph_op}
			\hat{A}_2&=
			\begin{tikzpicture}[baseline={([yshift=-.5ex]current bounding box.center)}]
				\node[circle,draw,minimum size=0.6cm] (1) at (0,1.8) {$1$};
				\node[circle,draw,minimum size=0.6cm] (2) at (0,0) {$2$}; 
				\node[fill, vertex] (v) at (3,1.8) { };
				\path[red, line width = 0.8pt] (1) edge (v);
				\path[blue, line width = 0.8pt] (2) edge (v);
				\node[fill, vertex] (u) at (3,0) { };
				\path[blue, line width = 0.8pt ] (1) edge  (u);
				\path[red, line width = 0.8pt] (2) edge (u);
			\end{tikzpicture} \nn \\
			&=
			\frac{1}{2}(\ha_{1,+} -\ha_{2,-})(\ha_{2,+} -\ha_{1,-}),
		\end{align} which corresponds to
		\begin{align}\label{optical_bell_sculpting}
			\frac{1}{2}(\ha_{1,H} -\ha_{2,V})(\ha_{2,H} -\ha_{1,V})
		\end{align} in our optical setup. 
	
	We implement two heralded subtraction operator of the form~\eqref{subtraction_opt} following the structure of the sculpting bigraph~\eqref{Bell} that corresponds to \eqref{optical_bell_sculpting}, which results in the following circuit: 
		\begin{align}
			\includegraphics[width=.45\textwidth]{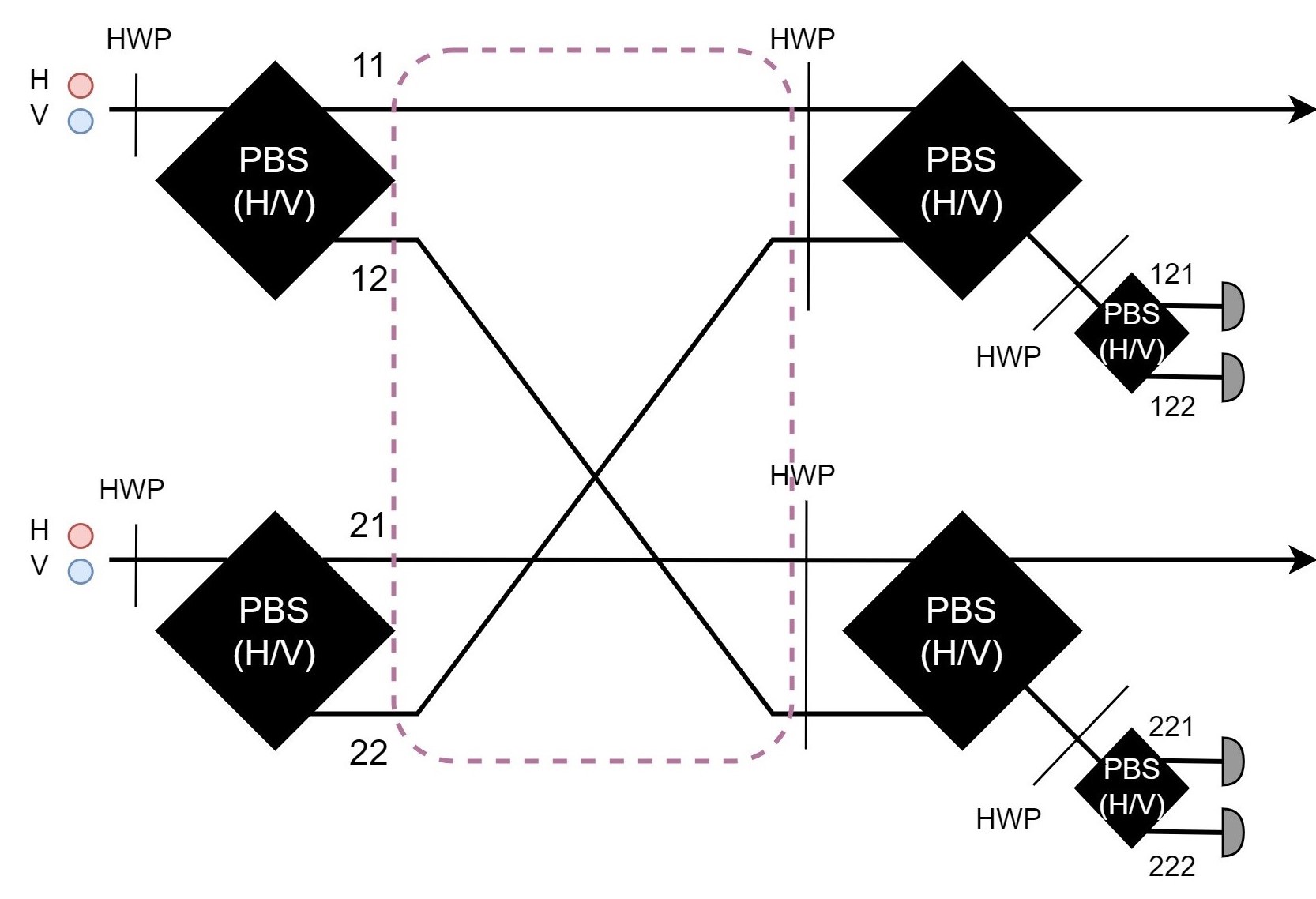}
		\end{align}
		By postselecting only the cases when each detector observes one photon, we generate Bell states as expected. Note that the wires of  heralded subtraction operators are attached as the structure of the Bell sculpting bigraph~\eqref{Bell} (see the wires in the dashed purple box). Appendix \ref{Bell_optics}  explains the state evolution in detail.
	
		In the case of W state and $N=3$  Type 5 state, we require subtraction operators for the superposition of larger than 2 spatial modes in the ancillae. This can be achieved by generalizing~\eqref{subtraction_opt}. Ref.~\cite{chin2023from}, Appendix A provides a more thorough analysis and optical circuits specific to these states. 
		It is worth noting that
		any bosonic system with linear operators that transforms the spatial and internal states as PBSs and HWPs can execute the same entanglement generation schemes as those given here.
	
		For the case of qudit entanglement generation with sculpting operators, we need a bosonic system that has a higher level of internal degree of freedom. For example, photons can have the orbital angular momentum (OAM) that can encode qudit information. Therefore, we can generate the GHZ qudit entanglement given in Sec.~\ref{qudit} with \emph{OAM beam splitters}~\cite{zou2005scheme} and \emph{OAM-only Fourier transformation operators}~\cite{kysela2020fourier}. OAM beam splitters  change the outgoing paths of photons with respect to the internal states, hence a $d$-level generalization of the PBS. OAM-only Fourier transformation operators transform the computational basis of the internal states, hence a $d$-level generalization of the HWP.

	\section{Discussions}\label{discussions}
	
	Our strategy for finding entanglement generation schemes based on linear bosonic systems with heralding  involves a two-step process: first, we find a theoretical sculpting operator that generates an entangled state. Second, we construct a concrete experimental circuit for such a sculpting operator. For the first step of finding sculpting operators, we have exploited graph techniques by imposing the correspondence relations of bosonic systems to bigraphs. We have shown that the graph picture of bosonic systems facilitates a powerful tool to find proper sculpting operators. For the second step, we have explained that a spatially overlapped subtraction operator can be installed in linear optical networks with heralding. 
		This operator allows us to design circuits for generating heralded entanglement by combining them based on sculpting bigraphs.
		As the simplest example, we have presented a Bell state circuit with the heralded subtraction operators, which can be extended to other sculpting schemes that we have proposed (see Ref.~\cite{chin2023from}). Our current results suggest several interesting future research directions.

		First, our formalism and strategy can be extended to encompass more complex qudit systems. We have introduced EPM bigraphs for finding qubit solutions, which can be generalized to qudit cases.We can suggest a more complete demonstration on the construction of desigining qudit heralded schemes with more solutions for qudit entanglement. We can encode such qudit entangled states as the orbital angular momentum (OAM) of photons with OAM beam splitters~\cite{zou2005scheme} and OAM-only Fourier transformation operators~\cite{kysela2020fourier}.
	
		Second, our sculpting protocol can identify other interesting multipartite entangled states. Since our graph approach provides a handy guideline to coming up with useful sculpting operators, we expect that it would be used to find heralded schemes for other crucial entanglements. 
		For example, Ref~\cite{chin2023linear} has recently found sculpting schemes for generating a special type of graph states, i.e., caterpillar graph states, with less photons than fusion gates~\cite{browne2005resource}. Therefore, one of our next goals will be to find, e.g., more general types of graph states, $N$-particle $N$-level singlet states~\cite{cabello2002n}, and cyclically symmetric states~\cite{Karczewski2019}.
	
	
	
	
	
	\section*{Data availability}
	
	All relevant data supporting the main conclusions are available upon reasonable request. Please refer to SC, sbthesy$@$gmail.com.
	
	\section*{Author contributions}
	SC is responsible for the methodology, solutions, and the manuscript. 
	MK was involved in theoretical discussions and supported writing. YSK worked on the optical circuit of the Bell state.

	\section*{Acknowledgements}
	SC is grateful to Prof. Ana Belen Sainz and Prof. Jung-Hoon Chun for their support on this research. 
	This research is funded by National
	Research Foundation of Korea (NRF, 2019R1I1A1A01059964, 2021M3H3A103657313, and 2022M3K4A1094774), Korea Institute of Science and Technology (2E31021), and Foundation for Polish Science (IRAP project, ICTQT, contract no.2018/MAB/5, co-financed by EU within Smart Growth Operational Programme).

	\section*{Competing interests}
	All authors declare no competing interests.

	\newpage	
	\onecolumngrid	
	\appendix

	\section{Directed bigraph mapping of many-boson systems with creation and annihilation operators}\label{comprehensive list}	
	
	In Section III of the main content, we have presented a list of correspondence relations between sculpting operators and bigraphs. Even if the list suffices to explain our sculpting protocol, we can propose a comprehensive $directed$ graph mapping of many-boson systems including both creation and annihilation operators. 
	
	In this directed graph mapping, both creation and annihilation operators correspond to unlabelled vertices. They are distinguished by the direction of edges attached to them.
	In other words, \emph{creation (annihilation) operators are denoted as unlabelled vertices whose edges go to (come from) circles.} The other correspondence relations are not changed from Table~I of the main content.  
	
	The directed graph mapping is useful to describe the thorough process of sculpting protocol. The initial state~\eqref{initial} is drawn in a directed graph as
	\begin{align}
		|Sym_N\>= \prod_{j=1}^N(\ha^\dagger_{j,0}\ha^\dagger_{j,1})|vac\>= 
		\begin{tikzpicture}[baseline={([yshift=-.5ex]current bounding box.center)}]
			\node[fill, vertex] (1) at (-2,3.8) { } ;
			\node[fill, vertex] (2) at (-2,3.2) { };
			\node[fill, vertex] (3) at (-2,2.3) { } ;
			\node[fill, vertex] (4) at (-2,1.7) { };   
			\draw[dashed, thick] (-2,1.3) -- (-2,.6); 
			\node[fill, vertex] (5) at (-2,.3) { } ;
			\node[fill, vertex] (6) at (-2,-.3) { }; 
			\node[circle,draw,minimum size=0.8cm] (X) at (0,3.5) {1};
			\node[circle,draw,minimum size=0.8cm] (Y) at (0,2) {2};
			\draw[dashed, thick] (0,1.3) -- (0,.7); 	
			\node[circle,draw,minimum size=0.8cm] (W) at (0,0) {N};  
			\path[line width = 0.8pt, ->] (1) edge (X);
			\path[dotted, line width = 0.8pt, ->] (2) edge (X);  
			\path[line width = 0.8pt, ->] (3) edge (Y);
			\path[dotted, line width = 0.8pt, ->] (4) edge (Y);  
			\path[line width = 0.8pt, ->] (5) edge (W);
			\path[line width = 0.8pt,dotted, ->] (6) edge (W);   
		\end{tikzpicture}.
	\end{align}		
	For example, the $N=2$ Bell state generating 
	scheme with a sculpting bigraph~\eqref{Bell} is expressed as the following directed bigraph
	\begin{align}
		|\P\>_{fin} =\hat{A}_2|Sym_2\> = (\hat{A}^{(1)}\hat{A}^{(2)})\ha^\dagger_{1,0}\ha^\dagger_{1,1}\ha^\dagger_{2,1}\ha^\dagger_{2,1}|vac\> 
		=
		\begin{tikzpicture}[baseline={([yshift=-.5ex]current bounding box.center)}]
			\node[circle,draw,minimum size=0.5cm] (1) at (0,1.5) {$1$};
			\node[circle,draw,minimum size=0.5cm] (2) at (0,0) {$2$}; 
			\node[fill, vertex] (v) at (1.5,1.5) { };
			\path[line width = 0.8pt,->, red] (1) edge (v);
			\path[line width = 0.8pt,->, blue] (2) edge (v);
			\node[fill, vertex] (u) at (1.5,0) { };
			\path[line width = 0.8pt,->, blue] (1) edge  (u);
			\path[line width = 0.8pt,->, red] (2) edge(u);
			\node[fill, vertex] (10) at (-1.5,1.8) { };
			\node[fill, vertex] (11) at (-1.5,1.2) { };
			\node[fill, vertex] (20) at (-1.5,.3) { };
			\node[fill, vertex] (21) at (-1.5,-.3) { };		
			\path[line width = 1pt, -> ] (10) edge  (1);
			\path[line width = 1pt,dotted, ->] (11) edge (1);
			\path[line width = 1pt, -> ] (20) edge  (2);
			\path[line width = 1pt, dotted, -> ] (21) edge (2);			
		\end{tikzpicture}. 
	\end{align}
	
	The directed bigraphs present a clear diagrammatic understanding of the identities~\eqref{qubit_identities} and~\eqref{qubit_identities_2}, which are translated into directed bigraphs respectively as
	\begin{align}
		\begin{tikzpicture}[baseline={([yshift=-.5ex]current bounding box.center)}]
			\node[circle,draw,minimum size=0.7cm] (j) at (0,1.5) {$j$};
			\node[fill, vertex,] (1) at (-1.8,.8) { };
			\node[fill, vertex] (2) at (-1.8,2.2) { };  
			\node[fill, vertex] (v) at (1.8,1.5) { };
			\path[line width = 0.8pt, color=red,->] (j) edge  (v);
			\path[line width = 0.8pt,->] (1) edge  (j); 
			\path[line width = 0.8pt, dotted, ->] (2) edge  (j);  
		\end{tikzpicture}~=~
		\begin{tikzpicture}[baseline={([yshift=-.5ex]current bounding box.center)}]
			\node[circle,draw,minimum size=0.7cm] (j) at (0,1.5) {$j$};
			\node[fill, vertex] (v) at (-1.8,1.5) { };
			\path[line width = 0.8pt, color=red,->] (v) edge  (j);
		\end{tikzpicture},~~~~
		\begin{tikzpicture}[baseline={([yshift=-.5ex]current bounding box.center)}]
			\node[circle,draw,minimum size=0.7cm] (j) at (0,1.5) {$j$};
			\node[fill, vertex,] (1) at (-1.8,.8) { };
			\node[fill, vertex] (2) at (-1.8,2.2) { };  
			\node[fill, vertex] (v) at (1.8,1.5) { };
			\path[line width = 0.8pt, color=blue,->] (j) edge  (v);
			\path[line width = 0.8pt,->] (2) edge  (j); 
			\path[line width = 0.8pt, dotted, ->] (1) edge  (j); ;  
		\end{tikzpicture}~=~-
		\begin{tikzpicture}[baseline={([yshift=-.5ex]current bounding box.center)}]
			\node[circle,draw,minimum size=0.7cm] (j) at (0,1.5) {$j$};
			\node[fill, vertex] (v) at (-1.8,1.5) { };
			\path[line width = 0.8pt, color=blue,->] (v) edge  (j);
		\end{tikzpicture}  
	\end{align}
	and 
	\begin{align}\label{path_identity_directed}
		&\begin{tikzpicture}[baseline={([yshift=-.5ex]current bounding box.center)}]
			\node[circle,draw,minimum size=0.7cm] (j) at (0,1.5) {$j$};
			\node[fill, vertex,] (1) at (-1.8,.8) { };
			\node[fill, vertex] (2) at (-1.8,2.2) { }; 
			\node[fill, vertex] (v) at (1.8,.8) { };
			\node[fill, vertex] (u) at (1.8,2.2) { };
			\path[line width = 0.8pt,->] (2) edge  (j); 
			\path[line width = 0.8pt, dotted, ->] (1) edge  (j);    
			\path[line width = 0.8pt, color=red, ->] (j) edge   (u);
			\path[line width = 0.8pt, color=blue, ->] (j) edge  (v);
		\end{tikzpicture}~=~0,~~\nn \\
		\phantom{sdfdf} \nn \\
		&\begin{tikzpicture}[baseline={([yshift=-.5ex]current bounding box.center)}]
			\node[circle,draw,minimum size=0.7cm] (j) at (0,1.5) {$j$};
			\node[fill, vertex,] (1) at (-1.8,.8) { };
			\node[fill, vertex] (2) at (-1.8,2.2) { }; 	
			\node[fill, vertex] (v) at (1.8,.8) { };
			\node[fill, vertex] (w) at (1.8,2.2) { };
			\draw[thick, dashed] (.7,1.25) to[bend right=50] (.7,1.75);	
			\path[line width = 0.8pt,->] (2) edge  (j); 
			\path[line width = 0.8pt, dotted, ->] (1) edge  (j); 
			\path[line width = 0.8pt, ->] (j) edge  (v);
			\path[line width = 0.8pt, ->] (j) edge  (w);
			\node at (1.4,1.5) {$n~(\geq 2)$} ;
		\end{tikzpicture}~=~		\begin{tikzpicture}[baseline={([yshift=-.5ex]current bounding box.center)}]
			\node[circle,draw,minimum size=0.7cm] (j) at (0,1.5) {$j$};
			\node[fill, vertex,] (1) at (-1.8,.8) { };
			\node[fill, vertex] (2) at (-1.8,2.2) { }; 		
			\node[fill, vertex] (v) at (1.8,.8) { };
			\node[fill, vertex] (w) at (1.8,2.2) { };
			\draw[thick, dashed] (.7,1.25) to[bend right=50] (.7,1.75);	
			\path[line width = 0.8pt,->] (2) edge  (j); 
			\path[line width = 0.8pt, dotted, ->] (1) edge  (j);  
			\path[dotted, line width = 0.8pt, ->] (j) edge  (v);
			\path[dotted, line width = 0.8pt, ->] (j) edge  (w);
			\node at (1.4,1.5) {$n~(\geq 2)$} ;
		\end{tikzpicture}~=~ 0. 
	\end{align} 
	By omitting the left part of circles, the above relations become the bigraph identities~\eqref{path_identity} of the main content.

	\section{Sculpting-operator-finding strategy}\label{strategy}
	
	In this section, we briefly explain the advantage of Property 1 that links the final entangled state and perfect matchings (PMs) of sculpting bigraphs. Even if we have found sculpting operators that generates entanglement by combining it with Property 2 in the main content, there are several reasons that Property 1 itself provides useful insight to analyze sculpting operators that do not correspond to effective PM diagrams.
	
	First, for a given bigraph that corresponds to a sculpting operator, we can immediately read the possible final state from the PMs of the bigraph. For the $N=2$ example, we can expect from~\eqref{N=2_A_expansion} that $\hat{A}_2$ has the potential to generate the Bell state before fixing the amplitudes. 
	
	Second,  we can apply all the PM diagram techniques developed in Ref.~\cite{chin2021graph} to our system. Since the final states in both approaches correspond to the summation of PMs in a given bigraph, necessary conditions for a bigraph to carry genuine entanglement in LQNs (see Theorem 1 of Ref.~\cite{chin2021graph}) are also valid to our protocol. 
	
	Third, in the same context as the second reason, we can consider bigraphs that generate entanglement in LQNs~\cite{chin2021graph} as strong candidates for sculpting operators that generate entanglement in our protocol.

	Based on these advantages, we can build a strategy to find sculpting operator for a genuinely entangled state~\cite{walter2016multipartite}.
	\begin{tcolorbox}[enhanced jigsaw,colback=black!10!white,boxrule=0pt,arc=0pt]
		\textbf{Sculpting-operator-finding strategy} 
		\begin{enumerate}
			\item Write down all the states that consist of the entangled state that we want to generate. Draw the PMs that correspond to the states. 
			\item  Draw a bigraph that has the above PMs. We choose a bigraph with mininal edges so that it has minimal collective paths.
			\item  Examine whether we can set the edge weights so that only PMs among the collective paths contribute to the final state.
			\item If we can find such an edge weight solution, it corresponds to the sculpting operators that generates the entangled state we expect. If we cannot, we try other bigraph with the same PMs but more edges.   
		\end{enumerate}
	\end{tcolorbox}
	In Step 2, we can use, e.g., a method suggested in Ref.~\cite{chin2021graph}, 3.2 to find bigraphs for a specific set of PMs. Note that Step 2 provides a significant benefit since reducing edges in bigraphs means reducing the number of possible no-bunching restrictions that we have to consider. Furthermore, a sculpting operator found in that way  usually can be constructed more efficiently since a smaller number of edges implies a small amount of resource to create operator superpositions. 
	One can understand the edge number as the coherence number~\cite{chin2017coherence} of a quantum state, which is a coherence monotone that quantifies the amount of coherence in a quantum system. Therefore, we can consider in a general sense that a system corresponding to a bigraph with more edges needs more quantum resource.

	\section{Qudit GHZ state}\label{qudit_GHZ}
	
	To show that $\hat{S}_{GHZ_{N,d}}|Sym_{N,d}\>$ constructs the GHZ state, we use the following identities:
	\begin{align}\label{identity_qudit}
		& (\ha_{\tu})^l(\ha_{\widetilde{d-1}})^{d-1-l}\prod_{s=0}^{d-1}\ha^\dagger_{s} = (-1)^{(d-1-l)} \frac{l!(d-1-l)!}{\sqrt{d}^{d-2}}\ha^\dagger_{\widetilde{d-1-l}}~~~(l\in \{ 0,1,\cdots ,d-1 \}), \nn \\
		&    (\ha_{\tu})^m(\ha_{\widetilde{d-1}})^{d-m}\prod_{s=0}^{d-1}\ha^\dagger_{s} = 0~~~ (m \in \{1,\cdots ,d-1 \}). 
	\end{align}
	In the above equations,
	we can check that the second identity is directly obtained by taking $\ha_\tu$ or $\ha_{\widetilde{d-1}}$ to the first identity. 
	
	Then the sculpting operator~\eqref{GHZ_d} is expanded as
	\begin{align}
		\hat{S}_{GHZ_d} &=\Big(\frac{1}{\sqrt{2}}\Big)^{(d-1)N}(\ha_{1,\tu}-\ha_{2,\widetilde{d-1}})^{d-1} (\ha_{2,\tu}-\ha_{3,\widetilde{d-1}})^{d-1} \cdots (\ha_{N,\tu}-\ha_{1,\widetilde{d-1}})^{d-1} \nn \\
		&= \Big(\frac{1}{\sqrt{2}}\Big)^{(d-1)N}\Big[\sum_{l_1=0}^{d-1}\binom{d-1}{l_1}(\ha_{1,\tu})^{l_1}(\ha_{2,\td})^{d-1-l_1} \Big] \cdots \Big[\sum_{l_N=0}^{d-1}\binom{d-1}{_N}(\ha_{N,\tu})^{l_N}(\ha_{0,\td})^{d-1-l_N} \Big].
	\end{align}	
	Then, using identities~\eqref{identity_qudit}, we have
	\begin{align}
		\hat{S}_{GHZ_d}|Sym_d\> &= (\frac{1}{\sqrt{2}})^{(d-1)N}\sum_{l} \Big[\binom{d-1}{l}\Big]^N(-1)^{N(d-1-l)}\prod_{k=1}^N\Big(\ha_{k,\tu}^l\ha_{k,\widetilde{d-1}}^{d-1-l}(\prod_{s=0}^{d-1}\ha^\dagger_{k,s})\Big) \nn \\
		&= \Big(\frac{(d-1)!}{\sqrt{2}^{d-1}\sqrt{d}^{d-2}}\Big)^N(|\tilde{0},\tilde{0},\cdots, \tilde{0} \>+ \cdots +|\widetilde{d-1},\widetilde{d-1},\cdots, \widetilde{d-1}\>),
	\end{align} i.e., 
	the qudit GHZ state in the quantum Fourier transformed basis.

	\section{Bell state generation in linear optics}\label{Bell_optics}
	
	\begin{center}
		\begin{figure}[t]
			\includegraphics[width=12cm]{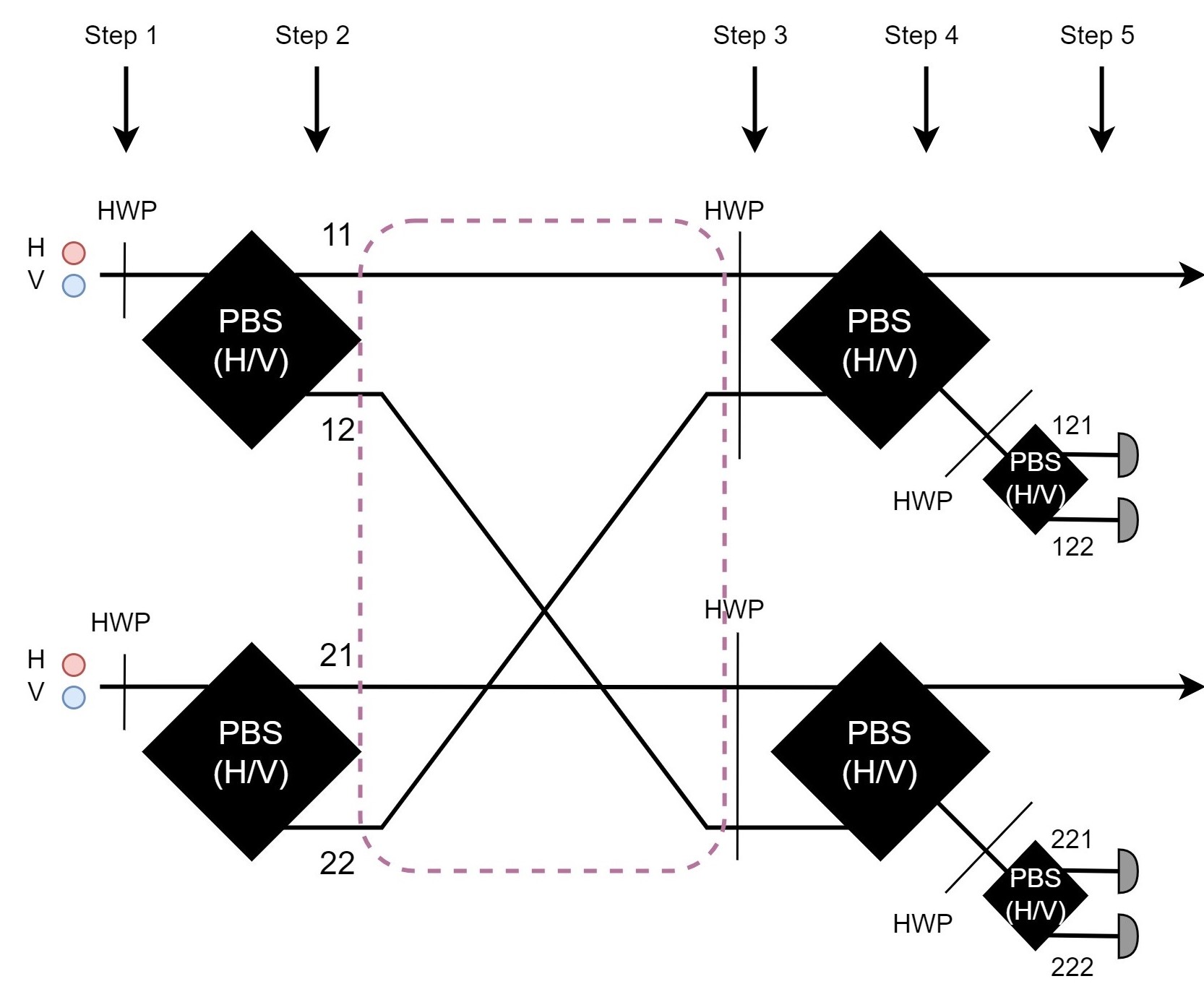}
			\caption{Linear optical circuit for generating the Bell state by 5 steps}
			\label{fig_bell2}
		\end{figure}
	\end{center}
	
	In this section, we explain the photon state evolution of our experimental scheme presented in Sec.~\ref{experiments}. The state evolution consists of 5 steps as denoted in Fig.~\ref{fig_bell2}. 
	\begin{itemize}
		\item Step 1
		\begin{align}
			\ha^\dagger_{1,H}\ha^\dagger_{1,V}\ha^\dagger_{2,H}\ha^\dagger_{2,V}|vac\>   ~~\to~~     
			\ha^\dagger_{1,D}\ha^\dagger_{1,A}\ha^\dagger_{2,D}\ha^\dagger_{2,A}|vac\>        
		\end{align}
		\item Step 2 
		\begin{align}
			&\frac{1}{4}(\ha_{11,H}^\dagger + \ha_{12,V}^\dagger)(\ha_{11,H}^\dagger - \ha_{12,V}^\dagger)(\ha_{21,H}^\dagger + \ha_{22,V}^\dagger)(\ha_{21,H}^\dagger - \ha_{22,V}^\dagger)|vac\> \nn \\
			&= \frac{1}{4}(\ha_{11,H}^{\dagger 2}  - \ha_{12,V}^{\dagger 2})(\ha_{21,H}^{\dagger 2}- \ha_{22,V}^{\dagger 2})|vac\>
		\end{align}
		
		\item Step 3 
		\begin{align}
			\frac{1}{4}(\ha_{11,D}^{\dagger 2} - \ha_{22,A}^{\dagger 2})(\ha_{21,D}^{\dagger 2} - \ha_{12,A}^{\dagger 2})|vac\>
		\end{align}
		\item Step 4
		\begin{align}
			&\frac{1}{16} \Big( (\ha^{\dagger}_{11,H} +\ha^{\dagger}_{12,V})^2 - (\ha^\dagger_{22,H} -\ha^\dagger_{21,V})^2 \Big)
			\Big( (\ha^\dagger_{21,H} +\ha^\dagger_{22,V})^2 - (\ha^\dagger_{12,H} -\ha^\dagger_{11,V})^2 \Big)|vac\> \nn \\
			&= \frac{1}{16} \Big( \ha^{\dagger~~2}_{11,H} +2\ha^{\dagger}_{11,H}\ha^{\dagger}_{12,V} +\ha^{\dagger~~2}_{12,V} - \ha^{\dagger~~2}_{22,H} +2\ha^{\dagger}_{22,H}\ha^{\dagger}_{21,V} -\ha^{\dagger~~2}_{21,V} \Big) \nn \\
			&~~~~\times
			\Big( \ha^{\dagger~~2}_{21,H} +2\ha^{\dagger}_{21,H}\ha^{\dagger}_{22,V} +\ha^{\dagger~~2}_{22,V} - \ha^{\dagger~~2}_{12,H} +2\ha^\dagger_{12,H}\ha^\dagger_{11,V} -\ha^{\dagger~~2}_{11,V} \Big)|vac\>
			` \end{align}
		\item Step 5
		\begin{align}
			\frac{1}{4}\Big( \ha^\dagger_{11,H}\ha^\dagger_{21,H}(\ha^\dagger_{121,H}-\ha^\dagger_{122,V})(\ha^\dagger_{221,H}-\ha^\dagger_{222,V}) + \ha^\dagger_{11,V}\ha^\dagger_{12,V}(\ha^\dagger_{121,H} + \ha^\dagger_{122,V})(\ha^\dagger_{221,H} + \ha^\dagger_{222,V})\Big)|vac\> 
		\end{align}
	\end{itemize}
	At Step 5, we sort out the states that click one of the two detectors in each PBS. When the two heralding detectors receive photons of the same polarization, we obtain a Bell state $\frac{1}{\sqrt{2}}(|HH\> + |VV\>)$. When the two heralding detectors receive photons of the opposite polarization, we obtain a different Bell state $\frac{1}{\sqrt{2}}(|HH\> - |VV\>)$.
	
	


	\twocolumngrid
	
	\bibliographystyle{unsrt}
	\bibliography{graphapproach}
	
\end{document}